\documentclass[11pt,a4paper]{article}
\pdfoutput=1
\usepackage{jheppub}
\usepackage{rotating}
\usepackage{array}
\usepackage{amsmath}
\usepackage[normalem]{ulem}
\usepackage{slashed}
\usepackage{booktabs}
\usepackage[pdftex,table,dvipsnames]{xcolor}
\usepackage{units}
\usepackage{multirow}
\usepackage{url}
\usepackage{parskip}
\usepackage[all]{nowidow}
\usepackage{placeins}

\usepackage{xspace}
\usepackage{subfig}
\usepackage{color}

\usepackage{dsfont}
\usepackage{soul}
\usepackage{hyperref}

\allowdisplaybreaks

\def \belletwo {Belle\,II\xspace}
\def \belle {Belle\xspace}
\def \lhcb {LHCb\xspace}
\def \babar {BaBar\xspace}

\newcommand{\Br}{\ensuremath{\mathcal{B}}\xspace}
\newcommand{\mb}{\ensuremath{\hat{M}_B^2}\xspace}
\newcommand{\ptk}{\ensuremath{p_T(K)}\xspace}
\newcommand{\phike}{\ensuremath{\phi_{Kp_{\rm miss}}}\xspace}
\newcommand{\me}{\ensuremath{\slashed{E}}\xspace}
\newcommand{\RomanNumeralCaps}[1]{\MakeUppercase{\romannumeral #1}}

\def\invab {\ensuremath{\mbox{\,ab}^{-1}}\xspace}
\def\bka {\ensuremath{B^+\to K^+ a}\xspace}
\def\uubar {\ensuremath{u\bar{u}}\xspace}
\def\ddbar {\ensuremath{d\bar{d}}\xspace}
\def\ssbar {\ensuremath{s\bar{s}}\xspace}
\def\ccbar {\ensuremath{c\bar{c}}\xspace}
\def\uubarg {\ensuremath{u\bar{u}(\gamma)}\xspace}
\def\ddbarg {\ensuremath{d\bar{d}(\gamma)}\xspace}
\def\ssbarg {\ensuremath{s\bar{s}(\gamma)}\xspace}
\def\ccbarg {\ensuremath{c\bar{c}(\gamma)}\xspace}
\def\tautaubar {\ensuremath{\tau\bar{\tau}}\xspace}
\def\charged {\ensuremath{B^+B^-}\xspace}
\def\mixed {\ensuremath{B^0\bar{B^0}}\xspace}
\def\btokinv{\ensuremath{B^+ \to K^+ a, a\to \me}\xspace}
\def\nbb{\ensuremath{N_{B\bar{B}}}\xspace}

\def\evtgen {\textsc{\hbox{EvtGen}}\xspace}
\def\kkmc {\textsc{\hbox{KKMC}}\xspace}
\def\pythia {\textsc{\hbox{PYTHIA8.2}}\xspace}
\def\tauola {\textsc{\hbox{TAUOLA}}\xspace}

\preprint{P3H-22-005,\,Nikhef 2022-001}

\title{Displaced or invisible? ALPs from $B$ decays at \belletwo}

\author[1]{T.~Ferber,}
\author[2]{A.~Filimonova,}
\author[3]{R.~Sch\"afer,}
\author[3]{and S.~Westhoff\,}

\affiliation[1]{Institute for Experimental Particle Physics,  Karlsruhe Institute of Technology (KIT), 76131 Karls\-ruhe, Germany}
\affiliation[2]{Nikhef, Science Park 105,1098 XG Amsterdam, The Netherlands}
\affiliation[3]{Institute for Theoretical Physics, Heidelberg University, 69120 Heidelberg, Germany\\}

\emailAdd{torben.ferber@kit.edu}
\emailAdd{a.filimonova@nikhef.nl}
\emailAdd{r.schaefer@thphys.uni-heidelberg.de}
\emailAdd{westhoff@thphys.uni-heidelberg.de}

\abstract{At colliders, neutral long-lived particles can be detected through displaced decay products or as missing energy. Which search strategy is better depends on the particle's decay length just as on the detector properties. We investigate the complementarity of displaced and invisible signatures for the \belletwo experiment. Focusing on axion-like particles $a$ produced from meson decays, we present a new search strategy for two-body decays \btokinv with missing energy \me. With $50\,$ab$^{-1}$ of data, \belletwo can probe light invisible resonances with branching ratio $\mathcal{B}(B^+\to K^+ a) \gtrsim 10^{-7}$ and decay length $c\tau_a \gtrsim 1\,$m. For axion-like particles, we expect the sensitivity of $B^+ \to K^+ \slashed{E}$ to small couplings to improve by up to two orders of magnitude compared to previous searches at collider and fixed-target experiments. For sub-GeV particles, $B^+ \to K^+ \slashed{E}$ at \belletwo and searches at beam-dump experiments are most sensitive; for heavier particles, searches for displaced vertices at \belletwo, long-lived particle experiments at the LHC, and future fixed-target experiments can probe the smallest couplings.}

\begin{document}

\maketitle

\flushbottom

\section{Introduction}
\label{SEC:introduction}
The lifetime of a particle is a matter of fact; its detection is a matter of perspective. At colliders, a long-lived particle (LLP) may decay within the detector and leave traces of displaced decay products. Or it may traverse the detector and only be reconstructed as missing energy from the remainder of the event. In either case, the sensitivity depends on the particle's source and kinematics just as much as on the experimental setup and detector properties. To determine the discovery potential at an experiment, we have to consider case by case: Which search strategy is most sensitive - \emph{displaced or invisible}?

Particle physics experiments are sensitive to LLPs within a huge range of masses and lifetimes~\cite{Lanfranchi:2020crw,Agrawal:2021dbo}. Collider experiments with a small detector coverage around the collision point, like \lhcb~\cite{Borsato:2021aum} and FASER~\cite{FASER:2019aik}, mostly rely on displaced decays within the detector. Similarly, fixed-target experiments with a directional source are often designed to detect visible final states~\cite{Dobrich:2019dxc,Tsai:2019buq,NA64:2020qwq}, but can also be sensitive to signatures with missing energy~\cite{NA62:2020xlg,NA62:2021zjw}. At these experiments, the search strategy for LLPs is usually determined by the geometrical setup. On the other hand, collider experiments with a large detector coverage like ATLAS, CMS and \belletwo can perform searches for both displaced vertices \cite{Alimena:2019zri,Duerr:2019dmv,Duerr:2020muu} and for missing energy \cite{Beltran:2010ww,Goodman:2010ku,LHCDarkMatterWorkingGroup:2018ufk,Belle-II:2021rof}. Here the optimal search strategy is much less predictable, especially for particles with decay lengths that are comparable with the scales of the detector.

In this work we compare displaced and invisible signatures at the \belletwo experiment. For concreteness, we focus on axion-like particles (ALPs) which can be resonantly produced in $B$ meson decays. \belletwo is perfectly suited to search for ALPs $a$ in $B\to K a,\, a\to \ell^+\ell^-$ decays with displaced leptons, as well as in $B\to K \me$ decays with missing energy. This versatility is mostly due to the large angular detector coverage and an excellent reconstruction efficiency for charged particles. Previously, \babar has performed searches for light resonances $X$ in $B\to K X, X \to \ell^+\ell^-$ with displaced leptons~\cite{BaBar:2015jvu} and a search for a leptophilic scalar $\phi$ in $e^+e^-\to\tau^+\tau^-\phi, \phi\to \ell^+\ell^-$ with displaced leptons~\cite{BaBar:2020jma}. Belle has searched for long-lived heavy neutral leptons $\nu_h$ in $B\to X\ell\nu_h, \nu_h \to \ell^-\pi^+$ with displaced leptons and pions~\cite{Belle:2013ytx}. For \belletwo, new searches with visible final states from LLP decays produced in $B$ meson decays ~\cite{Filimonova:2019tuy,Kachanovich:2020yhi,Bertholet:2021hjl}, $\tau$ decays~\cite{Cheung:2021mol,Guadagnoli:2021fcj}, or directly from $e^+e^-$ collisions~\cite{Duerr:2019dmv, Duerr:2020muu} have been proposed.

Searches for $B\to K\me$ with missing energy, on the contrary, have so far only been performed in the context of neutrinos in $B\to K\nu\bar{\nu}$~\cite{BaBar:2013npw,Belle:2017oht,Belle-II:2021rof}. These analyses have been optimized for three-body decay kinematics, and some of them even reject two-body decays to suppress background. While the BaBar search for $B\to K\nu\nu$~\cite{BaBar:2013npw} provides a 
 new physics interpretation, it is not optimised for new physics searches. 
 We will see that the BaBar results explore only small regions of the ALP parameter space that is not covered by other searches. The benefits of a dedicated two-body analysis at Belle II for new physics searches have been stressed in Ref.~\cite{MartinCamalich:2020dfe}.

We propose a new dedicated search strategy for invisible ALPs in two-body decays \btokinv at \belletwo. To optimize the sensitivity, we perform a detailed analysis of the dominant background processes and identify selection criteria for a low-background region. With such a search, \belletwo can probe ALPs with very small couplings that have not been explored in existing experiments. Our strategy can also be used more generally to search for light neutral scalars or pseudo-scalars produced from meson decays. To assess the complementarity of displaced and invisible searches, we compare our predictions for $B\to K \slashed{E}$ with existing predictions for $B\to K X, X \to \ell^+\ell^-$ decays at \belletwo~\cite{Filimonova:2019tuy}.

This article is organized as follows. In Sec.~\ref{SEC:alps}, we introduce a model for the production and decay of long-lived ALPs in meson decays. In Sec.~\ref{SEC:bounds}, we discuss existing bounds on the ALP parameter space from searches at flavor and fixed-target experiments. In Sec.~\ref{SEC:invisible}, we present our new search strategy for invisible ALPs with $B\to K \me$ decays at \belletwo and determine the projected sensitivity to light resonances. In Sec.~\ref{SEC:displaced}, we compare these projections with the search potential for displaced $B \to K a,\, a \to \ell^+\ell^-$ decays at \belletwo. We conclude in Sec.~\ref{SEC:conclusions} with an outlook on complementary ALP searches with displaced and invisible signatures at future experiments.

\section{ALP model and benchmarks}
\label{SEC:alps}
Axion-like particles are new pseudo-scalars whose interactions with Standard Model (SM) particles preserve a global shift symmetry $a \to a + c$, where $a$ is the ALP field and $c$ is a constant. Originally, axions were proposed as a solution to the strong CP problem in QCD~\cite{Peccei:1977hh,Peccei:1977ur,Weinberg:1977ma,Wilczek:1977pj}. More generally, axion-like particles can be predicted in many theories with spontaneously broken symmetries as pseudo Nambu-Goldstone bosons. At colliders, light pseudo-scalars can be produced in rare meson decays $M_1\to M_2\, a$~\cite{Berezhiani:1989fs,Bobeth:2001sq,Hiller:2004ii,Andreas:2010ms,Freytsis:2009ct,Batell:2009jf,Dolan:2014ska,Izaguirre:2016dfi,Gavela:2019wzg,Bauer:2021mvw}. Meson decays via flavor-changing neutral currents are suppressed in the Standard Model, but can be strongly enhanced if the ALP is resonantly produced. In this work, we are mostly concerned with $B\to K a$ decays which probe ALPs with masses $m_a < m_B - m_K$. At energy scales $\mu$ above the weak scale, $\mu_w$, the relevant ALP couplings are described by an effective Lagrangian~\cite{Georgi:1986df}\footnote{We adopt the notation from Ref.~\cite{Bauer:2020jbp}, but express the fermion fields above and below the weak scale in terms of their mass eigenstates $f$.}
\begin{align}\label{eq:lagrangian}
    \mathcal{L}_{\rm eff}(\mu > \mu_w) & = \sum_{i\neq j} \frac{c_{ij}^V(\mu)}{2}\,\frac{\partial^{\mu} a}{f_a} (\bar{f}_i \gamma_{\mu} f_j) + \sum_{i,j} \frac{c_{ij}^A(\mu)}{2}\,\frac{\partial^{\mu} a}{f_a} (\bar{f}_i \gamma_{\mu} \gamma_5 f_j)\\\nonumber
      & \quad + c_{GG}(\mu)\frac{a}{f_a}\,\frac{\alpha_s}{4\pi}\,G_{\mu\nu} \widetilde{G}^{\mu\nu} + c_{WW}(\mu)\frac{a}{f_a}\,\frac{\alpha_2}{4\pi}\,W_{\mu\nu} \widetilde{W}^{\mu\nu} + c_{BB}(\mu)\frac{a}{f_a}\,\frac{\alpha}{4\pi}\,B_{\mu\nu} \widetilde{B}^{\mu\nu}\,.
\end{align}
The ALP can be interpreted as the Goldstone boson of a spontaneously broken global chiral symmetry, broken at the cutoff scale $\Lambda = 4\pi f_a$ of the effective theory. In QCD, $f_a$ is closely related to the axion decay constant. In our analysis we set $f_a = 1\,$TeV.
Furthermore, $c_{ij}^V$ and $c_{ij}^A$ denote the ALP coupling to SM fermions $f_i,f_j$ from generations $i,j=1,2,3$ in the mass basis.   In general, the flavor and chiral structure of ALP couplings to fermions is arbitrary, with the only restriction that flavor-diagonal vector couplings $c_{ii}^V$ are absent due to the shift symmetry. Once a UV completion of the ALP is specified, the flavor structure of the ALP couplings is fixed. In this work, we assume that the fermion couplings are flavor-diagonal and flavor-universal at the cutoff scale, setting
\begin{align}
c_{ff}(\Lambda) \equiv c_{11}^A(\Lambda) = c_{22}^A(\Lambda) = c_{33}^A(\Lambda)\,,\qquad 
c_{ij}^A(\Lambda) = c_{ij}^V(\Lambda) = 0\ \,(i\neq j).
\end{align}
Finally, $c_{VV}$ with $V = \{G,W,B\}$ is the ALP coupling to gauge bosons with field strength tensor $V_{\mu\nu}$ and dual $\widetilde{V}_{\mu\nu} = \frac{1}{2}\epsilon_{\mu\nu\rho\sigma}V^{\rho\sigma}$, normalized to the respective gauge couplings. The sum over gauge indices is implicit.

Below the weak scale, the relevant terms of the effective Lagrangian read
\begin{align}
    \mathcal{L}_{\rm eff}(\mu < \mu_w) & = C_{dd'}(\mu)\,\frac{\partial^{\mu} a}{f_a} (\bar{d}_L \gamma_{\mu} d'_L) + h.c. + \sum_{f\neq t} \frac{c_{ff}(\mu)}{2}\,\frac{\partial^{\mu} a}{f_a} (\bar{f} \gamma_{\mu} \gamma_5 f)\\\nonumber
    & \quad + c_{\gamma \gamma}(\mu)\frac{a}{f_a}\,\frac{\alpha}{4\pi}
   \,F_{\mu\nu}\widetilde{F}^{\mu\nu} + c_{GG}(\mu)\frac{a}{f_a}\,\frac{\alpha_s}{4\pi}\,G_{\mu\nu} \widetilde{G}^{\mu\nu}.
\end{align}
Here $\alpha$ is the fine-structure constant and $c_{\gamma\gamma} = c_{WW} + c_{BB}$ is the ALP coupling to photons with field strength tensor $F_{\mu\nu}$.
 The Wilson coefficient $C_{dd'}(\mu)$ describes the effective ALP interaction with left-handed down-type quarks $dd' = \{sb,ds\}$ below the weak scale, generated through loops of virtual top quarks and electroweak bosons. At the weak scale, it is well approximated by~\cite{Bauer:2020jbp}
\begin{align}\label{eq:csb-coupling}
C_{dd'}(\mu_w) = V_{td}^\ast V_{td'}
 \Bigg[ & \big(1 - R_t(\mu_w,\Lambda)\big)\,c_{tt}(\Lambda)\,\frac{\alpha_t}{4\pi}\left(\frac{1}{2}\ln\frac{\mu_w^2}{m_t^2} - \frac{1}{4} - \frac{3}{2}\frac{1 - x_t + \ln x_t}{(1-x_t)^2}\right)\\\nonumber
 & + \frac{1}{9}R_t(\mu_w,\Lambda)\,c_{tt}(\Lambda)
 - c_{WW}\,\frac{\alpha_t}{4\pi}\frac{3\alpha}{2\pi s_w^2}\frac{1 - x_t + x_t \ln x_t}{(1-x_t)^2} \Bigg],
\end{align}

where $\alpha_t = y_t^2/4\pi$ and $x_t = m_t^2/m_W^2$. The function
\begin{align}
    R_t(\mu_w,\Lambda) \approx \frac{9}{2}\,\frac{\alpha_t(\mu_w)}{\alpha_s(\mu_w)} \Bigg[1 - \left(\frac{\alpha_s(\Lambda)}{\alpha_s(\mu_w)}\right)^{\frac{1}{7}}\Bigg]
\end{align} describes the RG evolution of the ALP coupling to top quarks, $c_{tt}$, from the cutoff scale $\Lambda$ down to the weak scale $\mu_w$. Due to our assumption of flavor-universal ALP couplings $c_{ff}$, see \eqref{eq:lagrangian}, we identify $c_{tt}(\Lambda) = c_{ff}(\Lambda)$ in what follows. Below the weak scale, the RG evolution of $C_{dd'}(\mu)$ is moderate, so that $C_{sb}(m_b)\approx C_{sb}(\mu_w)$ and $C_{ds}(m_s)\approx C_{ds}(\mu_w)$.\footnote{In our numerical analysis we include RG effects below the weak scale.} Due to our assumption of flavor-universal ALP-fermion couplings, the Wilson coefficients are related by $C_{sb} = V_{ts}^\ast V_{tb}/(V_{td}^\ast V_{ts})\,C_{ds}$. In scenarios with flavor-changing ALP-fermion couplings at the scale $\Lambda$, $s-b$ and $d-s$ couplings are in general independent parameters.

To make predictions for flavor observables, we evolve the coupling $c_{ff}(\Lambda)$ in \eqref{eq:lagrangian} down to the bottom mass scale $\mu_b = m_b$ via the renormalization group (RG)~\cite{Choi:2017gpf,MartinCamalich:2020dfe,Chala:2020wvs,Bauer:2020jbp}.\footnote{In general, the ALP couplings of left- and right-handed fermions evolve differently, with small effects on the flavor-diagonal couplings. For simplicity, we express the ALP-fermion couplings in terms of mass eigenstates $f$ in \eqref{eq:lagrangian}, but include the full evolution in our numerical analysis.} To perform the RG evolution and matching at the weak scale $\mu_w = m_Z$, we have used a code~\cite{bruggisser-grabitz} that is based on the results of Ref.~\cite{Bauer:2020jbp}. The gauge couplings $c_{VV}$ are not renormalized and therefore constant above the weak scale.

The production rate for an ALP from $B\to Ka$ decays is
    \begin{align}\label{eq:b-to-ka}
        \Gamma_{B\to K a} &= \frac{\pi}{4}\frac{C_{sb}^2(m_b)}{\Lambda^2}\, f_0^2\left(m_a^2\right)\,m_B\left(1-\frac{m_K^2}{m_B^2}\right)^2\lambda^{1/2}(m_B^2,m_K^2,m_a^2)\,,
    \end{align}
with the kinematic function $\lambda(a,b,c) =a^2+b^2+c^2 -2(ab + ac + bc)$. The scalar form factor $f_0\left(m_a^2\right)$ parametrizes hadronic effects in $B\to K$ transitions at
momentum transfer $q^2=m_a^2$. We implement the predictions for the scalar form factor $f_0\left(m_a^2\right)$ from Ref.~\cite{Gubernari:2018wyi}.

Expressed in terms of the three parameters $c_{ff}$, $c_{WW}$ and $m_a$, the branching ratio for $B^+\to K^+ a$ from \eqref{eq:b-to-ka} reads
\begin{align}\label{eq:b-to-ka-num}
    \mathcal{B}(B^+\to K^+ a) & = 0.1\left(\frac{c_{ff}(\Lambda)}{f_a\,[\text{TeV}]} - 0.0016\, \frac{c_{WW}(\Lambda)}{f_a\,[\text{TeV}]}\right)^2\frac{f_0^2(m_a^2)}{f_0^2(0)}\frac{\lambda^{1/2}(m_B^2,m_K^2,m_a^2)}{m_B^2-m_K^2}\,,
\end{align}
using $f_0(0)=0.329$~\cite{Gubernari:2018wyi}. The decay modes of the ALP vary strongly with the ALP mass. For $m_a < 2 m_e$, the ALP can only decay into two photons at a rate
\begin{align}\label{eq:ALPs-to-photons}
        \Gamma_{a\to\gamma\gamma} &= \frac{\alpha^2}{4\pi}\left|C_{\gamma\gamma}^{\text{eff}}(m_a)\right|^2 \frac{m_a^3}{\Lambda^2}\,,
\end{align}
with the effective coupling
\begin{align}
\label{alp-photon_effective-coupling}
    C_{\gamma\gamma}^{\rm eff}(\mu) \approx c_{WW}(\mu) + \mathcal{O}\Big(\frac{\alpha}{4\pi}\,c_{ff}\Big).
\end{align}
For $2 m_e < m_a < 3 m_\pi$, the ALP decays mostly into pairs of electrons or muons~\cite{Bauer:2017ris,Bauer:2020jbp},
\begin{align}\label{eq:ALPs-to-leptons}
        \Gamma_{a\to\ell\bar{\ell}} &= 2\pi m_a\frac{\left|C_{\ell\ell}^{\rm eff}(m_a)\right|^2 m_\ell^2}{\Lambda^2}\,\sqrt{1-\frac{4m_\ell^2}{m_a^2}}\,,\qquad \ell = \{e,\mu\}\,,
\end{align}
with the effective coupling\,\footnote{Even if fermion couplings are absent at the cutoff scale, $c_{ff}(\Lambda) = 0$, the ALP can still decay into leptons through loop effects of the gauge couplings. These decays are strongly suppressed by the electromagnetic coupling, they are not relevant for our analysis.}
\begin{align}
    C_{\ell\ell}^{\rm eff}(\mu) = c_{ff}(\mu) + \mathcal{O}\Big(\frac{\alpha^2}{16\pi^2}\, c_{WW}\Big).
\end{align}
For masses $m_a > 3 m_\pi$, the ALP can decay into hadrons. Below the scale of QCD confinement, the rate can be calculated in chiral perturbation theory; at high energies a perturbative treatment is possible. In the intermediate range around $\mu \approx 1\,\text{GeV}$, predictions of hadronic ALP decays are affected by large hadronic uncertainties and should be used with caution. For more details on our treatment of hadronic decays and a compendium of analytic expressions for all partial decay widths, we refer to App.~\ref{app:decay-widths}.

Besides decays into SM particles, ALPs may also have exotic decay channels. In particular, an ALP could serve as a mediator to a dark sector and decay into invisible final states at a rate $\Gamma_{a\to \rm inv}$. 
Assuming that no other decay channels exist, the total width of the ALP is given by
\begin{align}\label{eq:decay-width}
    \Gamma_a = \Gamma_{a\to \gamma\gamma} + \sum_{\ell = e,\mu,\tau}\Gamma_{a\to \ell\bar{\ell}}\, \Theta(m_a - 2 m_\ell) + \Gamma_{a\to \text{had}}\,\Theta(m_a - 3m_{\pi}) + \Gamma_{a\to \rm inv}\,.
\end{align}
The decay width of the ALP determines the lifetime $\tau_a = \Gamma_a^{-1}$. In our analysis, we focus on ALP decays into leptons and photons. Hadronic decays enter the phenomenology only through the lifetime.

\begin{figure}[t!]
    \centering
    \includegraphics[width=0.49\textwidth]{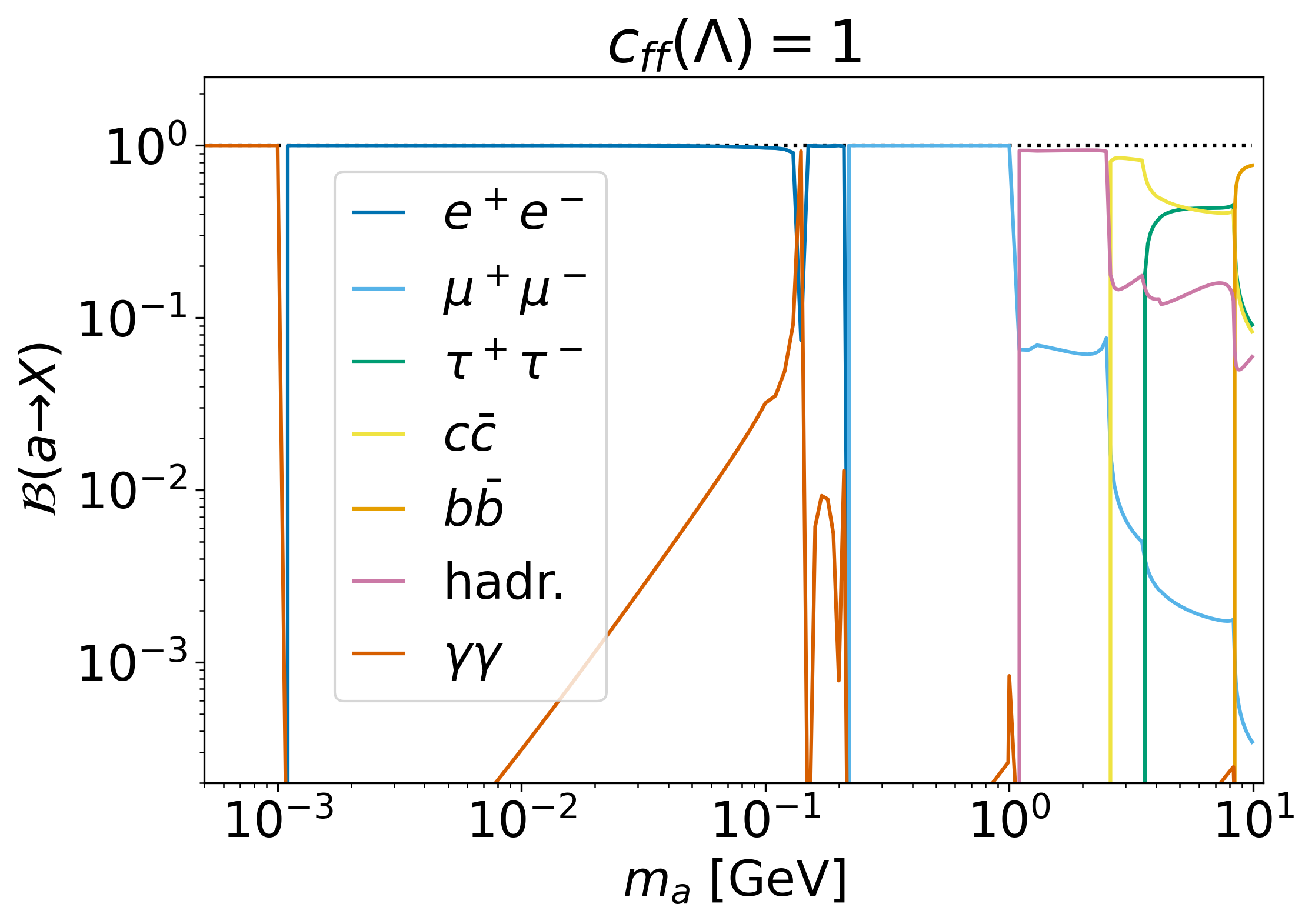}\hfill
    \includegraphics[width=0.49\textwidth]{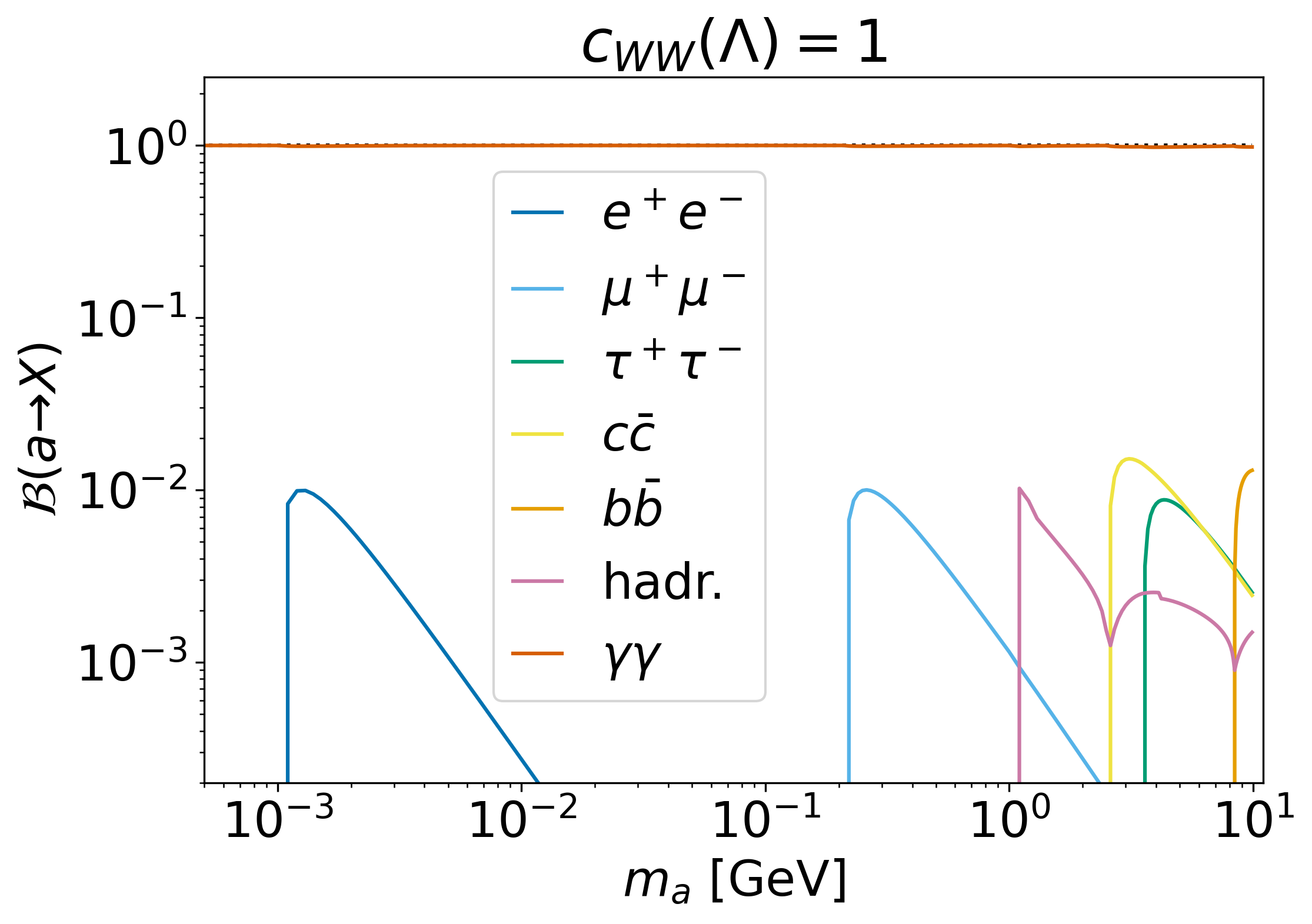}
    \caption{Branching ratios for ALP decays $a\to X$ in the $c_{ff}$ scenario (left) and the $c_{WW}$ scenario (right). The ALP couplings are fixed to $c_{ff}(\Lambda) = 1$ and $c_{WW}(\Lambda) = 1$ at the cutoff scale $\Lambda = 4\pi\,$TeV.}
    \label{fig:branching ratios}
\end{figure}

For our analysis we define two benchmark scenarios:
\begin{align}\label{eq:benchmarks}
    \text{``$c_{ff}$''}:\qquad & \text{ALP coupling to fermions} & c_{ff}(\Lambda)\neq 0,\ c_{WW}(\Lambda) = 0\,,\\\nonumber
    \text{``$c_{WW}$''}:\qquad & \text{ALP coupling to gauge bosons} & c_{WW}(\Lambda) \neq 0,\ c_{ff}(\Lambda) = 0\,.
\end{align}
The corresponding branching ratios of the ALP into the various final states are shown in Fig.~\ref{fig:branching ratios}. ALP decays to invisible final states are absent in these scenarios, so that $\Gamma_{a\to \text{inv}} = 0$ in \eqref{eq:decay-width}.

In the $c_{ff}$ scenario, the partial decay rates of the ALP scale as
\begin{align}\label{eq:alp-decays-1}
    \Gamma_{a \to \ell\bar{\ell}}\propto c^2_{ff}(m_a)\,,\qquad \Gamma_{a\to \gamma\gamma} \propto \displaystyle{\frac{\alpha}{4\pi}} c_{ff}^2(m_a)\,.
\end{align}
For masses $2 m_e < m_a < 3 m_\pi$, the ALP decays mostly into leptons. Decays into photons are loop-suppressed and dominate only for $m_a < 2 m_e$. 

In the $c_{WW}$ scenario, the ALP decays according to
\begin{align}\label{eq:alp-decays-2}
    \Gamma_{a \to \ell\bar{\ell}}\propto \left(\frac{\alpha}{4\pi}\right)^4 c^2_{WW}(m_a),\qquad \Gamma_{a \to \gamma\gamma}\propto c^2_{WW}(m_a)\,.
\end{align}
Here the ALP decays mostly into photons, while decays into leptons are loop-suppressed.

\begin{figure}[t!]
  \centering
  \vspace*{0.5cm}
  \includegraphics[scale=0.65]{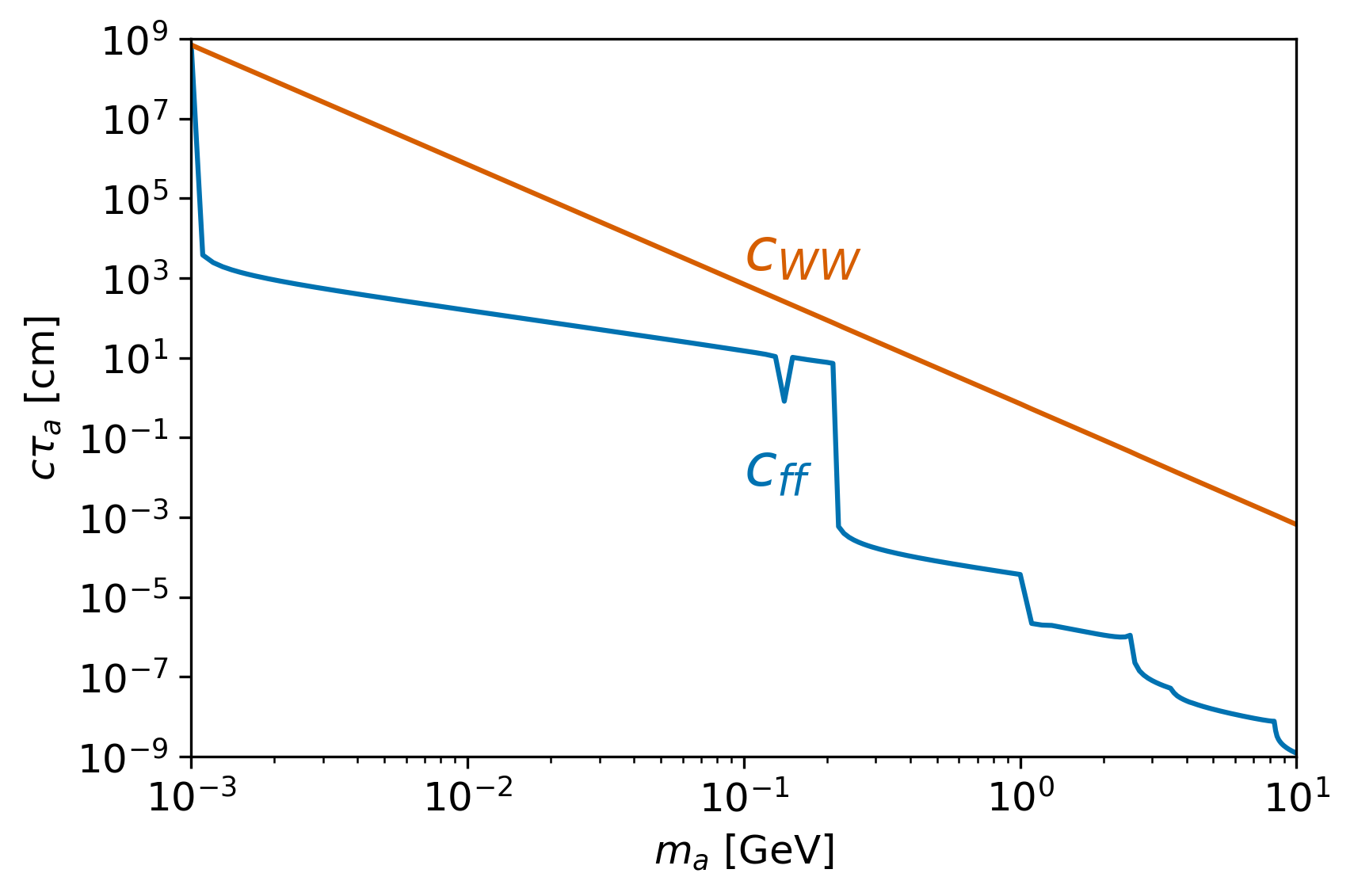}
  \caption{ALP decay length $c\tau_a$ as a function of the ALP mass $m_a$, for the two benchmark scenarios with $c_{ff}(\Lambda) = 1$ (blue) and $c_{WW}(\Lambda) = 1$ (orange) with $\Lambda = 4\pi\,$TeV.
    \label{fig:decay-length}}
\end{figure}

In Fig.~\ref{fig:decay-length}, we display the proper decay length $c\tau_a$ of the ALP as a function of its mass $m_a$ in the two benchmark scenarios with fixed couplings $c_{ff}(\Lambda) = 1$ and $c_{WW}(\Lambda) = 1$, respectively. In the $c_{WW}$ scenario, the lifetime of the ALP is larger because ALP decays through gauge couplings are rare compared to decays through fermion couplings. For small ALP masses and/or small couplings, the decay length is macroscopic and the ALP tends to decay at a distance from the production point.

\section{Bounds from flavor and fixed-target experiments}
\label{SEC:bounds}
The search strategy for ALPs at flavor experiments depends on the lifetime and decay modes of the ALP. At short lifetimes, the ALP decays within the detector and may be observed in $B\to K X$ or $K\to \pi X$ decays with detectable final states $X = \{\mu^+\mu^-,e^+ e^-,\gamma\gamma\}$. At long lifetimes, the ALP can be stable at detector scales and leave signatures with missing energy $\me$, such as $B\to K \me$ or $K\to \pi \me$. For ALPs with intermediate lifetimes, it is a priori not clear whether searches with visible or invisible final states are most sensitive.

The expected event rate in the detector scales with the decay probability of the ALP. The probability to find an ALP with boost factor $\beta\gamma$ at a distance $r$ from its production point is given by
\begin{align}
    \mathds{P}_a(r|\beta\gamma) = \exp\left(-\frac{r}{d_a}\right),\qquad d_a = \beta\gamma c\tau_a\,,
\end{align}
where $d_a$ is the ALP's decay length, defined in the laboratory frame. To obtain the average probability $\langle \mathds{P}_a \rangle$ for an ALP to decay \emph{outside} the detector in a sample of $N$ events, we sum over all trajectories inside the detector volume and take into account the respective boost, 
\begin{align}
\langle \mathds{P}_a \rangle = \frac{1}{N}\sum_{k=1}^N\, \mathds{P}_a(r_k|\beta\gamma_k)\,.
\end{align}
Here $r_k$ is the distance between the ALP production point and the point where it leaves the detector.  If the ALP is produced in meson decays, the expected event rate for ALP decays inside the detector is given by
\begin{align}\label{eq:alp-vis}
    N_a(B\to K X) = N_{B}\,\mathcal{B}(B\to K a)\times \mathcal{B}(a \to X) \times \big(1 - \langle \mathds{P}_a\rangle\big),
\end{align}
where $N_{B}$ is the number of $B$ mesons produced in an experiment. For invisible ALPs, the event rate is
\begin{align}\label{eq:alp-inv}
    N_a(B\to K \me) = N_{B}\,\mathcal{B}(B\to K a)\times \big( \mathcal{B}(a \to X)\times \langle\mathds{P}_a \rangle + \mathcal{B}(a \to \text{inv})\big).
\end{align}
Here $X$ refers to visible final states. Analogous expressions apply for $K\to \pi$ decays.

\begin{figure}[t!]
  \centering
  \includegraphics[scale=0.35]{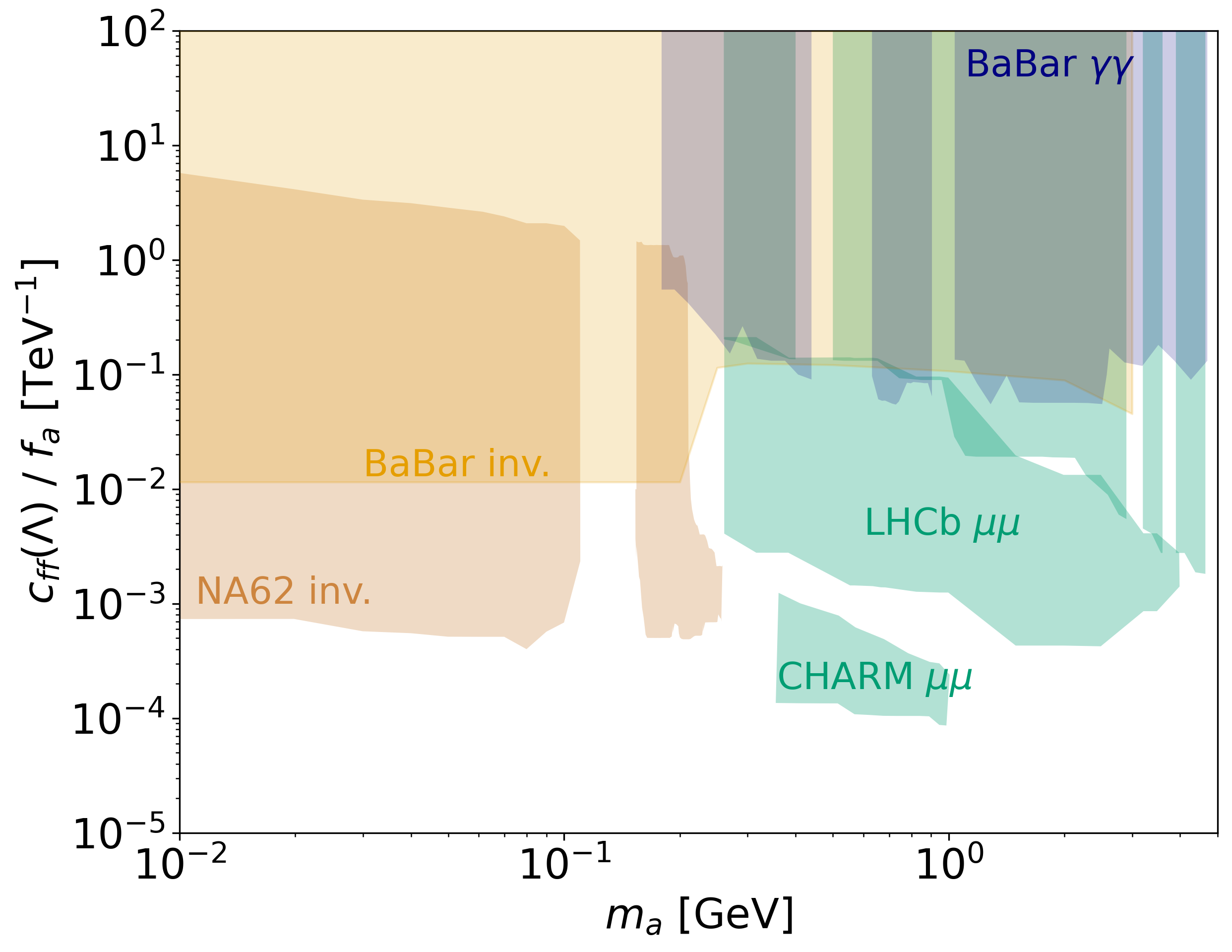}\hspace*{0.1cm} \includegraphics[scale=0.35]{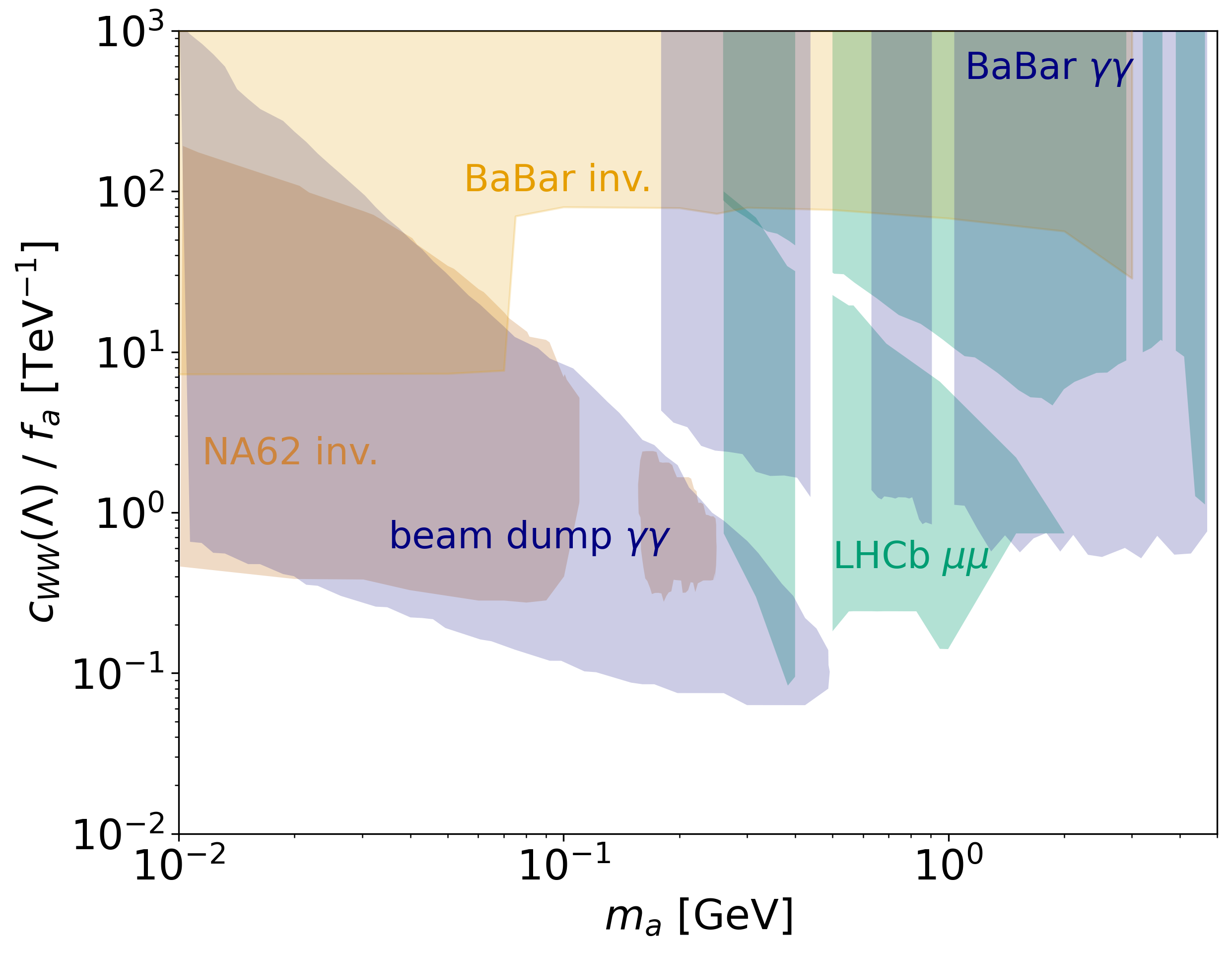}
  \caption{Bounds on the effective ALP coupling to fermions, $c_{ff}(\Lambda)/f_a$ (left), and weak gauge bosons, $c_{WW}(\Lambda)/f_a$ (right), as a function of the ALP mass $m_a$. Shown are bounds obtained from searches for $B^+\to K^+ \nu\bar{\nu}$ at 90\% CL by \babar~\cite{BaBar:2013npw} (yellow), $K^+\to \pi^+ X$ at 90\% CL by NA62~\cite{NA62:2020xlg,NA62:2021zjw} (orange), $B^+\to K^+ \gamma\gamma$ at 90\% by BaBar~\cite{BaBar:2021ich}, for di-photon signals at 90\% CL at beam-dump experiments~\cite{Dolan:2017osp}, notably at NuCal~\cite{Blumlein:1990ay}, CHARM~\cite{BERGSMA1985458} and E137~\cite{Bjorken:1988as}
  (blue), and searches for $B \to K X,\,X \to \mu^+\mu^-$ with prompt and displaced muons at 95\% CL by LHCb~\cite{LHCb:2015nkv,LHCb:2016awg} and CHARM~\cite{Dobrich:2018jyi} (green). Only one ALP coupling is present at $\Lambda = 4\pi\,$TeV; all other couplings are set to zero. Some of the bounds apply more generally in other scenarios, see the discussion in Sec.~\ref{SEC:generalization}. \label{fig:bounds}}
\end{figure}

In this section we investigate the interplay of visible and invisible ALPs at flavor and fixed-target experiments. We derive bounds from existing searches for rare meson decays that can be interpreted in terms of ALPs. In Sec.~\ref{sec:lepton-fs}, we focus on signatures with displaced leptons, most notably a search by \lhcb for light resonances in $B\to K \mu^+\mu^-$ decays with displaced muon pairs~\cite{LHCb:2016awg}. In Sec.~\ref{sec:photon-fs}, we discuss signatures with photons in the final state, in particular the recent search for ALPs in $B\to K\gamma\gamma$ by \babar~\cite{BaBar:2021ich}. In Sec.~\ref{sec:invisible-fs}, we investigate searches with missing energy. We reinterpret a search for $B\to K \nu\bar{\nu}$ decays by \babar~\cite{BaBar:2013npw} for invisible ALPs and derive bounds from a recent search for $K^+\to \pi^+X,\,X\to \text{inv}$ at NA62~\cite{NA62:2020xlg,NA62:2021zjw}. In Fig.~\ref{fig:bounds} we summarize the strongest bounds on ALPs in the two benchmark scenarios from~\eqref{eq:benchmarks}.
All bounds apply for the effective ALP couplings $c_{ff}(\Lambda)$ and $c_{WW}(\Lambda)$ at the cutoff scale $\Lambda = 4\pi\,\text{TeV}$. To derive these bounds from the measured observables, we have RG-evolved the coupling $c_{ff}(\mu)$ in our predictions from $\mu = \Lambda$ down to the relevant scales for $B$ and $K$ decays, $\mu = m_b$ or $\mu = m_s$, as described in Sec.~\ref{SEC:alps}. The coupling $c_{WW} = c_{WW}(\Lambda)$ is constant above the weak scale and little affected by the RG evolution at lower scales.

\subsection{ALP decays to leptons}\label{sec:lepton-fs}
ALPs that decay into leptons $\ell = \{e,\mu\}$ within the detector volume can be probed in $B \to K \ell^+\ell^-$ and $K \to \pi \ell^+\ell^-$ decays. The \lhcb collaboration has performed searches for a light resonance $X$ in $B \to K X,\,X \to \mu^+\mu^-$ decays with prompt~\cite{LHCb:2015nkv} and displaced~\cite{LHCb:2016awg} muon pairs, as well as an inclusive search for di-muon vertices~\cite{LHCb:2020ysn}. We reinterpret the search with displaced muons~\cite{LHCb:2016awg}, which is most sensitive to ALPs with long decay lengths. Based on 3\,fb$^{-1}$ of LHC data, \lhcb has derived upper limits on the product of branching ratios, $\mathcal{B}(B^+\to K^+X)\mathcal{B}(X \to \mu^+\mu^-)$. The results are reported for di-muon invariant masses in the range $250\,\text{MeV} < m_{\mu\mu} < 4700\,\text{MeV}$ and lifetimes of $X$ ranging from 0.1 to 1000\,ps. Our reinterpretation of these limits for ALPs with masses $m_a = m_{\mu\mu}$ is shown in Fig.~\ref{fig:bounds}. At the upper end of the excluded mass region (lower green area), $m_a \approx 1\,\text{GeV}$, the decay into muons is strongly suppressed by hadronic decays and the search loses sensitivity. The bounds on the coupling in the $c_{ff}$ scenario (left panel) are stronger than in the $c_{WW}$ scenario (right panel), due to the larger production rate \eqref{eq:b-to-ka-num} and branching ratio~\eqref{eq:alp-decays-2} into muons.

Larger ALP couplings are excluded by a similar search with less displaced di-muons (upper green area)~\cite{LHCb:2015nkv}, which is sensitive to ALPs with lifetimes $\tau_a < 0.2\,$s. A previous inclusive search for displaced vertices by \babar~\cite{BaBar:2015jvu} is not competitive with \lhcb in the considered mass range~\cite{Filimonova:2019tuy}.

Small couplings can also be constrained with the CHARM experiment. In Ref.~\cite{Dobrich:2018jyi}, the authors have reinterpreted a search for sterile neutrinos at CHARM~\cite{BERGSMA1985458}. They report model-independent bounds on $\mathcal{B}(B^+\to K^+X)\mathcal{B}(X \to \mu^+\mu^-)$ as a function of the lifetime $\tau_X$ for four fixed ALP masses. We interpolate between the provided mass benchmarks and show the resulting excluded regions in the ALP parameter space in Fig.~\ref{fig:bounds} (green). The search is only sensitive to the $c_{ff}$ scenario, where the ALP-lepton coupling is generated at tree level. In principle, CHARM could also search for lighter ALPs in $K\to \pi a,a\to \ell^+\ell^-$ decays. However, as explained in Ref.~\cite{Winkler:2018qyg}, most kaons are absorbed in the target, which drastically reduces the sensitivity to ALP decays in the far detector. We do not expect bounds from CHARM beyond the reach of missing energy searches at NA62 and BaBar, see Sec.~\ref{sec:invisible-fs}.

ALPs with masses below the di-muon threshold can be probed with rare kaon decays. The NA48/2 experiment at CERN has measured the rare decays $K^+\to \pi^+ \mu^+ \mu^-$~\cite{NA482:2010zrc} and $K^+\to \pi^+ e^+ e^-$~\cite{NA482:2009pfe}. The measurement focuses on the SM topology, selecting three-track vertices from kaon decays. Due to this selection criterion the measurement cannot be reinterpreted for ALPs with long decay lengths, where the pion and di-lepton momenta do not point back to the same vertex. For ALPs with short decay lengths, the parameter region for $m_a > 2 m_\mu$ has been excluded by \lhcb~\cite{LHCb:2015nkv}. For lighter ALPs with large couplings, $K^+\to \pi^+ e^+ e^-$ can set bounds in the range $140\,\text{MeV} < m_a < 354\,\text{MeV}$. In Sec.~\ref{sec:invisible-fs}, we will see that this parameter region has been excluded by a recent search for $K^+\to \pi^+\slashed{E}$ at NA62~\cite{NA62:2020xlg}. We therefore do not attempt to reinterpret the $K^+\to \pi^+\ell^+\ell^-$ measurements.

\subsection{ALP decays to photons}\label{sec:photon-fs}
While ALP signatures with leptons mostly probe the fermion coupling $c_{ff}$, final-state photons are sensitive to the gauge coupling $c_{WW}$. Here we derive bounds on $c_{WW}$ from searches for ALPs with photons at colliders and fixed-target experiments; for previous similar analyses see for instance Refs.~\cite{Dolan:2017osp,Dobrich:2019dxc}.

Recently the \babar collaboration has performed a dedicated search for ALPs in $B^+\to K^+ \gamma\gamma$ decays~\cite{BaBar:2021ich}. They report direct bounds on $\mathcal{B}(B^+ \to K^+ a)\mathcal{B}(a\to \gamma\gamma)$ for ALPs with masses of $175\,\text{MeV} < m_a < 4.78\,$GeV and decay lengths $c\tau_a <1\,$mm, and for $175\,\text{MeV} < m_a < 2.5\,$GeV for longer decay lengths up to $c\tau_a = 100\,$mm. We show the resulting bounds in blue in Fig.~\ref{fig:bounds}.

In Fig.~\ref{fig:bounds}, we show the resulting bounds in the two ALP scenarios (blue area). In the $c_{WW}$ scenario, the ALP decays to photons at tree level and the search by \babar is very sensitive to small couplings. The bound is determined by the decreasing production rate, $\mathcal{B}(B^+ \to K^+ a) \sim c_{WW}^2$, except for small ALP masses, where the decay length exceeds $c\tau_a = 100\,\text{mm}$. In the $c_{ff}$ scenario, the ALP decay into photons is loop-induced. The branching ratio $\mathcal{B}(a\to \gamma\gamma)$ is strongly suppressed, see~\eqref{eq:alp-decays-1}, because the decay length of the ALP is dominated by the decay into fermions. The bounds on $c_{ff}$ are therefore relatively weak and the sensitivity is limited by the rate, rather than the lifetime, except for the smallest accessible ALP masses. Nevertheless, it is remarkable that a photon search is sensitive to ALP couplings to fermions through quantum effects alone.

Searches for rare kaon decays are sensitive to ALPs with smaller masses. The fixed-target experiment E949 has searched for $K^+\to \pi^+\gamma\gamma$ decays in a phase-space region corresponding to di-photon invariant masses $m_{\gamma\gamma} < 108\,$MeV~\cite{E949:2005qiy}. The search targets the three-body decay topology; kinematic selections have been applied to reduce background, which limit the sensitivity to two-body kaon decays. Due to the lack of information on these selections and knowing that NA62's $K^+\to \pi^+ \slashed{E}$ search (see Sec.~\ref{sec:invisible-fs}) is more sensitive to long-lived ALPs, we do not attempt to reinterpret the E949 analysis.

At $e^+ e^-$ colliders, searches for di-photon resonances from direct ALP production via $e^+e^- \to \gamma a$, $a\to \gamma\gamma$ are an interesting alternative to meson decays. The \belletwo collaboration has performed such a search and derived bounds on the ALP coupling to photons for masses $200\,\text{MeV} < m_a < 9.7\,$GeV~\cite{Belle-II:2020jti}. With the current sensitivity, the reach is comparable with similar searches at LEP II and ATLAS~\cite{Knapen:2016moh}, but is expected to improve with more data.

Long-baseline experiments are sensitive to ALPs with even smaller couplings, because the predicted production rate is high and the long baseline allows to probe long decay lengths. The currently strongest bounds on the ALP-photon coupling arise from a combination of the proton beam-dump experiments NuCal, CHARM and E137~\cite{Dobrich:2019dxc,Blumlein:1990ay,Blumlein:2011mv,BERGSMA1985458,Bjorken:1988as}. ALPs can be produced from bremsstrahlung photons\footnote{At proton beam dumps, another source of photons are meson decays.} or via Primakoff conversion in the target, $\gamma Z \to a Z$.

For NuCal, CHARM and E137, we interpret the bounds from Ref.~\cite{Dobrich:2019dxc} in the $c_{WW}$ scenario. The bound cannot be directly translated to the $c_{ff}$ scenario; however, due to loop-suppressed ALP decays to photons we expect a lower sensitivity than in missing energy searches at BaBar and NA62. In Fig.~\ref{fig:bounds}, we see that NuCal, CHARM and E137 probe ALPs with masses $m_a \lesssim 500\,$MeV; the lower cutoff at $m_a= 200\,$MeV was set by the experimental analysis. In the excluded region the decay length of the ALP matches the baseline and beam energy of the experiments. For the electron beam experiment NA64, we have checked that the bounds on $c_{WW}/f_a$ obtained for $m_a \lesssim 60\,$MeV by the collaboration in Ref.~\cite{NA64:2020qwq} do not exceed the reach of $B\to K\me$ at BaBar.

\subsection{Invisible ALP decays}\label{sec:invisible-fs}
ALPs that escape the detector can be caught in signatures with missing energy, notably in the rare meson decays $B\to K \slashed{E}$ and $K\to \pi \slashed{E}$.

Both \babar~\cite{BaBar:2013npw} and Belle~\cite{Belle:2017oht} have searched for the rare decays $B\to K^{(\ast)} \nu\bar{\nu}$ with SM neutrinos in the final state. The analysis by \belle has been optimized for three-body final states, explicitly removing events with two-body decay kinematics. The search is therefore insensitive to ALPs produced in $B\to K a$ decays.

BaBar's search for $B\to K \nu\bar{\nu}$~\cite{BaBar:2013npw} reports differential distributions of the momentum transfer $s_B = q^2/m_B^2$ in $B\to K$ transitions, which allows us to derive bounds on invisible ALPs produced via $B\to K a$. The analysis employs hadronic $B$ tagging, so that the $B\to K \me$ decay kinematics can be fully reconstructed. The results based on $\nbb = 471\times 10^6$ $B\bar{B}$ pairs produced at the $\Upsilon(4S)$ resonance are presented in bins of width $\Delta s_B = 0.1$ for $B \to K \me$ decays with charged and neutral kaons. We reinterpret the results from charged $B^+\to K^+$ decays, which yield the strongest bounds obtained from an individual channel.\footnote{In Ref.~\cite{MartinCamalich:2020dfe}, the analysis has been interpreted including charged and neutral kaons, which leads to a somewhat stronger bound compared with our result.} An ALP with mass $m_a$ would increase the event rate $\Delta \mathcal{B}$ in the bin containing $s_B = m_a^2/m_B^2$ by
\begin{align}
\Delta \mathcal{B}_a = \frac{\Delta N_a(B^+\to K^+ \me)}{0.514\,N_{B\bar{B}}}\,,
\end{align}
with $N_a(B\to K\me)$ from \eqref{eq:alp-inv}. The factor $0.514$ accounts for the different production rates of $B^+$ and $B^0$ mesons in the $B\bar{B}$ sample. We require the sum of ALPs and neutrinos to fall within 1.64\,$\sigma$ of the observed event rate in this bin,\footnote{We have verified that the detector acceptance for kaons from $B\to K \nu\bar{\nu}$ and from $B\to K a$ decays is nearly identical and does not affect the interpretation.}
  \begin{align}
        \Delta \mathcal{B}\left(B^+\to K^+\nu\bar\nu\right) + \Delta \mathcal{B}_a \le (1+1.64\sigma)\,\Delta\mathcal{B}\left(B^+\to K^+ \me\right)\,,
    \end{align}
which corresponds to a confidence level of 90\,\% assuming Gaussian uncertainties. The neutrino background is very small, $\Delta \mathcal{B}\left(B^+\to K^+\nu\bar\nu\right)=7.3\cdot10^{-7}$ for $s_B < 0.1$, which corresponds to $m_a \lesssim 1.6\,\text{GeV}$. The resulting bounds on our ALP benchmarks are shown in Fig.~\ref{fig:bounds} (yellow area).

The bounds on the ALP coupling in the $c_{ff}$ scenario are generally stronger than in the $c_{WW}$ scenario, due to the different production rates $\mathcal{B}(B\to K a)$, see~\eqref{eq:b-to-ka-num}. The variation of the bounds with the ALP mass $m_a$ is related to its decay length. For small masses, the ALP has a macroscopic decay length, see Fig.~\ref{fig:decay-length}, and decays mostly outside the detector. In this region the search for \btokinv is most effective. For larger masses, the ALP tends to decay close to its production point. Here the sensitivity is set by the detector acceptance, i.e., by the probability for the ALP decay products to miss the detector region. In either regime, the sensitivity is largely insensitive to the ALP mass. The small dip around $m_a \gtrsim 1.6\,$GeV is due to a slightly higher experimental sensitivity of \babar for $s_B > 0.1$.

The sensitivity jump between light and heavy ALPs is explained by the interplay of production rate and decay length: When decreasing the coupling $c/f_a$ of an ALP with a fixed mass, the production rate decreases as $\mathcal{B}(B\to Ka) \sim (c/f_a)^2$, while the probability not to decay inside the detector increases as $\exp(-(c/f_a)^{-2})$. The compensation of these two effects is very sensitive to the coupling, which explains the height of the sensitivity jump. In the $c_{ff}$ scenario, the jump occurs close to the muon threshold $m_a \approx 2 m_\mu$, where the ALP decay length decreases abruptly as the decay into muons opens. Searches for $K^+\to \pi^+ \slashed{E}$ decays are sensitive to invisible ALPs with masses $m_a < m_K - m_\pi$. We calculate the branching ratio for $K^+\to \pi^+ a$ in analogy to $B^+\to K^+ a$ in Sec.~\ref{SEC:alps} and obtain
\begin{align}
\mathcal{B}(K^+\to \pi^+ a) = 4.12\times 10^{-4}\left(\frac{c_{ff}(\Lambda)}{f_a\,[\text{TeV}]} - 0.0016\, \frac{c_{WW}(\Lambda)}{f_a\,[\text{TeV}]}\right)^2 \frac{\lambda^{1/2}(m_K^2,m_\pi^2,m_a^2)}{m_K^2 - m_\pi^2}\,,
\end{align}
where $c_{ff}(\Lambda) = c_{tt}(\Lambda)$, as in \eqref{eq:b-to-ka-num}. The scalar form factor $f_0^K(q^2)$ has been computed in lattice QCD~\cite{Carrasco:2016kpy}. For simplicity, we set $f_0^K(q^2) = 1$ in our analysis, which slightly overestimates the ALP production rate.

At the NA62 experiment, kaons are produced from a high-energetic proton beam impinging on a beryllium target. The decay products can be observed in a $60\,$m long fiducial volume. Recently the collaboration has interpreted their measurement of $K^+\to \pi^+\nu\bar{\nu}$ decays for two-body decays $K^+\to \pi^+X$ with a long-lived new particle $X$~\cite{NA62:2020xlg,NA62:2021zjw}. They report limits on the branching ratio $\mathcal{B}(K^+\to \pi^+ X)$ as a function of the $X$'s mass for lifetimes $\tau_X > 100\,$ps. We show the reinterpreted bounds in Fig.~\ref{fig:bounds} (orange area).  In both scenarios, $K^+\to \pi^+ \slashed{E}$ at NA62 is sensitive to smaller ALP couplings than $B^+\to K^+\nu\bar{\nu}$.

NA62's measurement of $K^+\to \pi^+ \nu\bar{\nu}$ is not sensitive to ALPs with masses $m_a \approx m_{\pi^0}$, due to a large background from $K^+\to \pi^+ \pi^0$ decays. This region has been explored by NA62 in a dedicated analysis of invisible $\pi^0$ decays, which has been reinterpreted for $K^+\to \pi^+ X$ decays with $\tau_X > 100\,$ps~\cite{NA62:2020pwi}. The resulting bounds are shown in Fig.~\ref{fig:bounds} (yellow).

The KOTO experiment has searched for invisible particles $X$ in $K_L^0 \to \pi^0 X$ decays~\cite{KOTO:2018dsc}. However, KOTO's bound on $\mathcal{B}(K_L^0 \to \pi^0 X)$ assumes that all particles $X$ are invisible in the experiment. An interpretation of the analysis for ALPs with a finite decay width would require a dedicated simulation and recast, which goes beyond the scope and purpose of this work.

Existing searches for visible and invisible ALP signatures complement each other in their sensitivity. For masses below the di-muon threshold, searches with missing energy at BaBar and NA62, as well as displaced photons at beam dumps lead the sensitivity to small ALP couplings. For heavier ALPs, searches for displaced di-muons at LHCb and CHARM, and a new search with displaced photons by \belletwo dominate the bounds. Taken together, current searches set mass-dependent upper bounds on the ALP couplings
\begin{align}\label{eq:current-bounds}
c_{ff}(\Lambda)/f_a \lesssim (10^{-3} - 10^{-2})\,\text{TeV}^{-1},\qquad c_{WW}(\Lambda)/f_a \lesssim (10^{-2} - 1)\,\text{TeV}^{-1}\,,
\end{align}
up to a few gaps in the parameter space, see Fig.~\ref{fig:bounds}.

\subsection{Generalization beyond the benchmark scenarios}\label{SEC:generalization}
The bounds discussed in this section and displayed in Fig.~\ref{fig:bounds} apply for two specific ALP scenarios. Some of these bounds can be translated to more general scenarios. ALP production at BaBar, LHCb and CHARM happens through $B$ decays; all other searches rely on kaon decays. For NA62~\cite{NA62:2020xlg,NA62:2021zjw}, the lower bounds on $c_{ff}$ and $c_{WW}$ apply generally to invisibly decaying ALPs with a non-zero $C_{ds}$ coupling. In such scenarios, the bounds on $c_{ff}$ or $c_{WW}$ can be translated into a bound on $C_{ds}$ using Eq.~\eqref{eq:csb-coupling}, which allows for a direct interpretation for ALPs with flavor-changing couplings at the scale $\Lambda$. Similarly, the bounds from the BaBar search for $B^+\to K^+ \nu\bar{\nu}$~\cite{BaBar:2013npw} at low (high) ALP masses apply for invisible (promptly decaying) ALPs. Also here, the bounds on $c_{ff}$ or $c_{WW}$ can be translated to $C_{sb}$ using Eq.~\eqref{eq:csb-coupling}. Similarly, the $c_{ff}$ bound from $B \to K X,\,X \to \mu^+\mu^-$ at LHCb~\cite{LHCb:2015nkv} and the $c_{WW}$ bound from $B^+\to K^+ \gamma\gamma$ at BaBar~\cite{BaBar:2021ich} can be translated into bounds on $C_{sb}$ for promptly decaying ALPs. For all remaining searches, the production and decay of the ALP are correlated when fixing $c_{ff}$ or $c_{WW}$, which prevents a direct interpretation of the results in other scenarios.

\subsection{Comments on bounds from cosmology and astrophysics}
Complementary bounds on ALP couplings can arise from astrophysical and cosmological searches. Most analyses derive constraints on the effective ALP-photon coupling $C_{\gamma\gamma}^{\rm eff}$, and therefore on $c_{WW}$, see~\eqref{alp-photon_effective-coupling}. In Fig.~\ref{fig:bounds}, right, bounds on ALP emission during the explosion of the supernova SN 1987A constrain the lower left corner of the parameter space for $m_a \lesssim 200\,$MeV~\cite{Masso:1995tw,Lee:2018lcj,Lucente:2020whw}, assuming ALP production through the Primakoff process in the core of the supernova. However, these bounds depend on the explosion mechanism and other astrophysical aspects, see for instance~\cite{Bar:2019ifz}.

Similar regions of parameter space are also constrained by modifications of the cosmic history in the presence of light ALPs~\cite{Millea:2015qra,Depta:2020wmr,Balazs:2022tjl}. Most of these bounds constrain the lifetime of the ALP; the translation to its couplings is model-dependent. Even smaller ALP-photon couplings, going beyond Fig.~\ref{fig:bounds}, right, are constrained by supernova bounds on ALP decays to photons~\cite{Jaeckel:2017tud}.

In a similar way, cosmology and astrophysics also constrain ALP couplings to matter particles, which translate into bounds on $c_{ff}$. For a recent summary on sub-MeV ALPs, see Ref.~\cite{Green:2021hjh}. At larger ALP masses, SN1987A constrains couplings to nucleons~\cite{Raffelt:2006cw} and to electrons~\cite{Lucente:2021hbp}, which translate into bounds of $c_{ff}$ in the lower left corner of Fig.~\ref{fig:bounds}, left.

Some of the bounds mentioned above are compiled in Ref.~\cite{Bauer:2021mvw} for ALPs with pure photon couplings or pure electron couplings. In the parameter region that can be probed at Belle II with missing energy searches, see Fig.~\ref{fig:projections_combined}, supernova bounds predominate. Since these bounds are affected by large uncertainties, as discussed above, we prefer not to include them in Fig.~\ref{fig:bounds} and Fig.~\ref{fig:projections_combined}.\\

\section{Search strategy for invisible ALPs at Belle II}
\label{SEC:invisible}
In what follows, we will explore the potential of \belletwo to improve the sensitivity to long-lived ALPs with searches for missing energy or displaced vertices.

The search for \btokinv is experimentally similar to the search for the SM process $B^+\to K^+\nu\bar{\nu}$, where \belletwo recently pioneered an inclusive tagging approach to increase the signal efficiency \cite{Belle-II:2021rof}. This approach is in contrast with the exclusive method, where one $B$ meson is completely reconstructed in a hadronic final state before reconstructing the signal-side $B$ meson in the rest of the event \cite{BaBar:2013npw}. In the text, we only refer to $B^+\to K^+ a$ decays; charge-conjugate channels are implied in what follows. In contrast to $B^+\to K^+\nu\bar{\nu}$, $B^+\to K^+ a$ is a two-body decay with very specific event kinematics. The signal hence consists of a single charged kaon that can be reconstructed in the tracking detectors, and missing momentum. Final states with neutral kaons could be included in a future analysis.

We study the \belletwo sensitivity for a dataset corresponding to an integrated luminosity of 0.5\,\invab, roughly equivalent to the full \babar dataset, and a luminosity of 50\,\invab, corresponding to the expected final dataset of \belletwo. We generate events using the \belletwo  Analysis  Software  Framework \cite{Kuhr:2018lps, the_belle_ii_collaboration_2021_5574116}. We do not simulate the detector response, but approximate efficiencies and acceptance as explained below. We expect the effect of such simplifications on triggers and resolution to be rather small. We also do not include effects of beam-induced backgrounds, which however will reduce the missing energy resolution in the real experiment.

\subsection{Signal and background simulation}
We generate the events in the $e^+e^-$ centre-of-mass frame with the nominal \belletwo collision energy of $\sqrt{s}= 10.58$\,GeV, then boost and rotate them to the \belletwo laboratory frame. The \belletwo beam parameters are $E$(e$^+$) = 4.002\,GeV and $E$(e$^-$) = 7.004\,GeV with a 41.5\,mrad crossing angle between the beams and the $z$-axis. In the laboratory frame the $z$-axis is oriented along the bisector of the angle between the direction of the electron beam and the reverse direction of the positron beam. All selections below refer to parameters in the lab frame unless noted otherwise.

To produce signal events we generate events for $\Upsilon(4S)\to B^+B^-$, followed by decays of one $B$ through $B^+\to K^+ a$ and the other $B$ decaying generically. The decays of the charged $B$ mesons are simulated using the \evtgen generator \cite{Lange:2001uf}: 
The signal-side $B$ meson decay is using the \evtgen phase space model PHSP, while the generic $B$ meson decay is using all available $B$ meson decay modes of the \belletwo decay descriptions. For each ALP mass benchmark we generate 10k events.\footnote{We have simulated 12 ALP benchmarks with the masses $m_a$= 5\,MeV, 50\,MeV, 70\,MeV, 100\,MeV, 200\,MeV, 250\,MeV, 300\,MeV, 500\,MeV, 1\,GeV, 2\,GeV, 3\,GeV, and 4\,GeV. For clarity, we only show 4 representative masses in Fig.~\ref{fig:signal-kinematics}, but we use all benchmark masses for constructing the bounds in Fig.~\ref{fig:results_Br} and~\ref{fig:results}.}

To produce background events we generate $\Upsilon(4S)\to \charged$ and $\Upsilon(4S)\to \mixed$ using \evtgen. In addition we generate continuum backgrounds from $e^+e^-\to \uubarg$, $\ddbarg$, $\ssbarg$, $\ccbarg$, using \kkmc\,\cite{Jadach:1999vf}, simulate the hadronization of the quarks with \pythia\,\cite{Sjostrand:2014zea}, and model the decays of generated mesons with \evtgen. We also include background from $e^+e^- \to \tau^+\tau^-$, using \kkmc and \tauola to generate and decay the tau leptons\,\cite{Jadach:1990mz}. For each background channel, we generate 10 million events and use the following cross sections\,\cite{Belle-II:2018jsg} to normalize their rates: 1.61\,nb (\uubar), 0.40\,nb (\ddbar), 0.38\,nb (\ssbar), 1.30\,nb (\ccbar), 0.919\,nb (\tautaubar), 0.54\,nb (\charged), and 0.51\,nb (\mixed).

For our analysis we only consider final-state charged particles ($e, \mu, \pi, K, p$) with transverse momenta $p_T>0.2$\,GeV and photons with energies $E>0.05$\,GeV. For both charged particles and photons, we require that they are in the acceptance of the central drift chamber of \belletwo with polar angles $17^{\circ} < \theta < 150^{\circ}$. Neutrons, $K^0_L$, and neutrinos are counted as invisible. Short-lived resonances like $K^0_S$ and $\pi^0$ are decayed promptly and are included via their final-state decay products. We do not simulate the detector response for the generated particles, but we include an 80\,\% kaon identification efficiency and a 5\,\% pion misidentification rate~\cite{Collaboration:2052} for both signal and background. In addition, we assume a track-finding efficiency of 99\,\%~\cite{Collaboration:2035} per charged particle, and we approximate the photon detection efficiency with 100\,\%. We assume normal distributions for the relative momentum resolution of charged particles, $\Delta p/p = 0.5\,\%$~\cite{BelleIITrackingGroup:2020hpx}, and for the relative energy resolutions for photons, $\Delta E/E = 5\,\%$\,\cite{Belle-II:2018jsg}.
 
\subsection{Event selection}
Signal $B$ meson candidates contain one reconstructed kaon and missing momentum. All remaining final-state charged particles and photons are hence associated with the decay of the tag-side $B$ meson. We reconstruct a missing-momentum vector $p_{\rm{miss}} = (E_{\rm miss},\vec{p}_{\rm miss})$, defined as the total momentum needed to balance the sum of momenta of all detected charged particles and photons, and the well-known initial state in the $e^+e^-$ collision.
 
For background rejection we construct the following three kinematic variables, defined in the lab frame:
\begin{itemize}
    \item \ptk is the transverse momentum of the kaon with the highest transverse momentum in the event, in the following called the \emph{leading kaon},
    \item $\mb=(E_{\rm{miss}}+E_K)^2 - (\vec{p}_{\rm{miss}}+\vec{p}_{K})^2$ is the reconstructed mass of the signal $B$ meson candidate, i.e., the invariant mass of the leading kaon and missing momenta,
    \item \phike is the angle between the leading kaon and the missing momentum vector, calculated using the four-vectors of all final-state particles and the incoming beams.
\end{itemize}

\begin{figure}[t!]
  \centering
  \includegraphics[width=\textwidth]{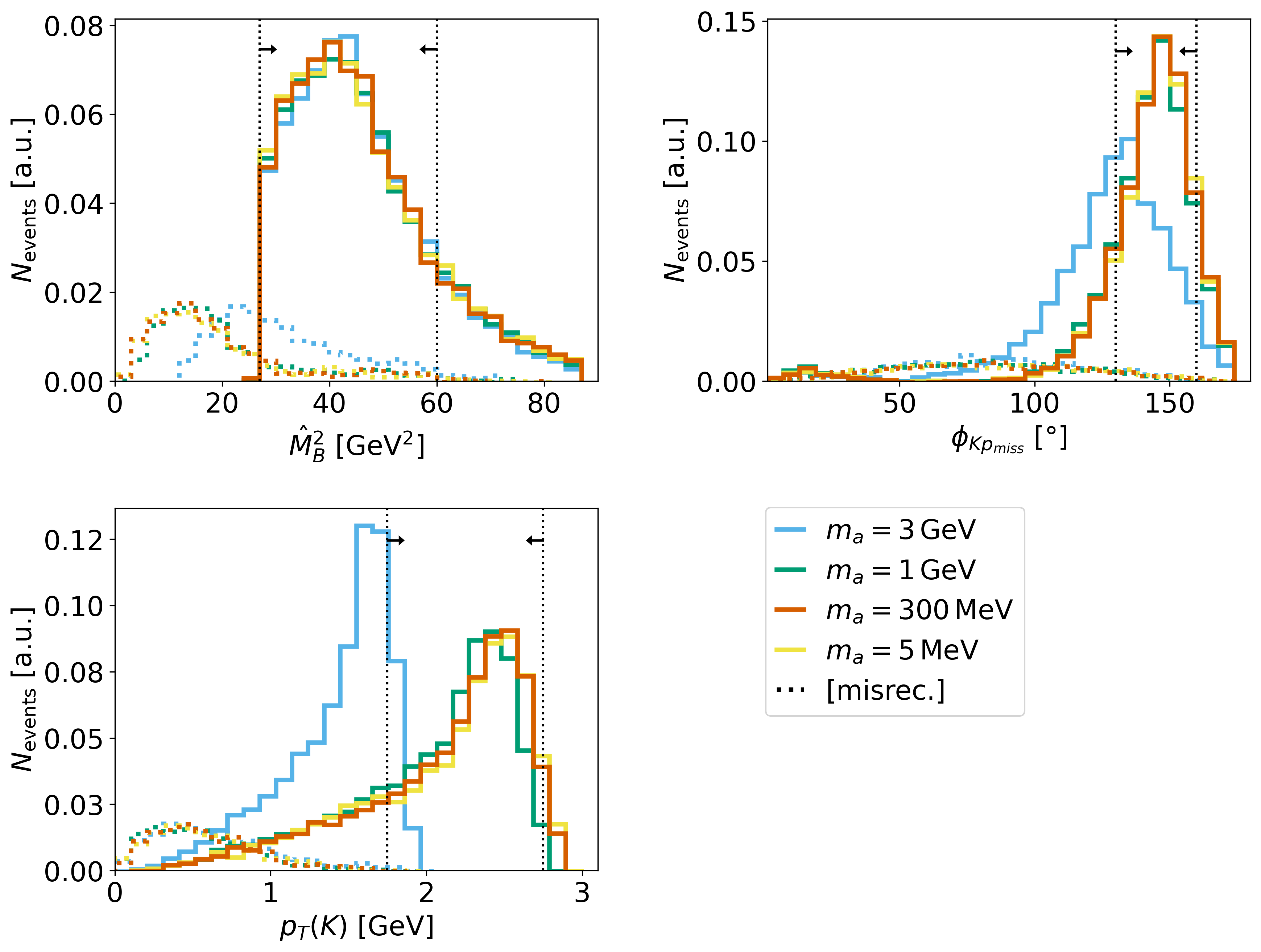}
  \caption{Kinematic distributions of ALPs produced from $B^+\to K^+ a$ for benchmarks of fixed ALP masses $m_a$. Top left: reconstructed $B$ meson mass, \mb; top right: opening angle of the leading kaon against the missing momentum, \phike; bottom left: transverse momentum of the leading kaon, \ptk. All variables are defined in the lab frame. Shown are signal events (solid curves) and misreconstructed signal events (dotted curves). The number of events is given in arbitrary units. The dotted vertical lines indicate selection cuts; the arrows point toward the signal region.
    \label{fig:signal-kinematics}}
\end{figure}

Choosing the leading kaon for \ptk ensures that only one signal candidate per event exists, but also introduces background from wrongly reconstructed signal candidates with kaons that belong to the tag-side $B$ meson. We label these events as [\emph{misrec.}] when we illustrate our results.

\begin{figure}[t!]
  \centering
  \includegraphics[width=\textwidth]{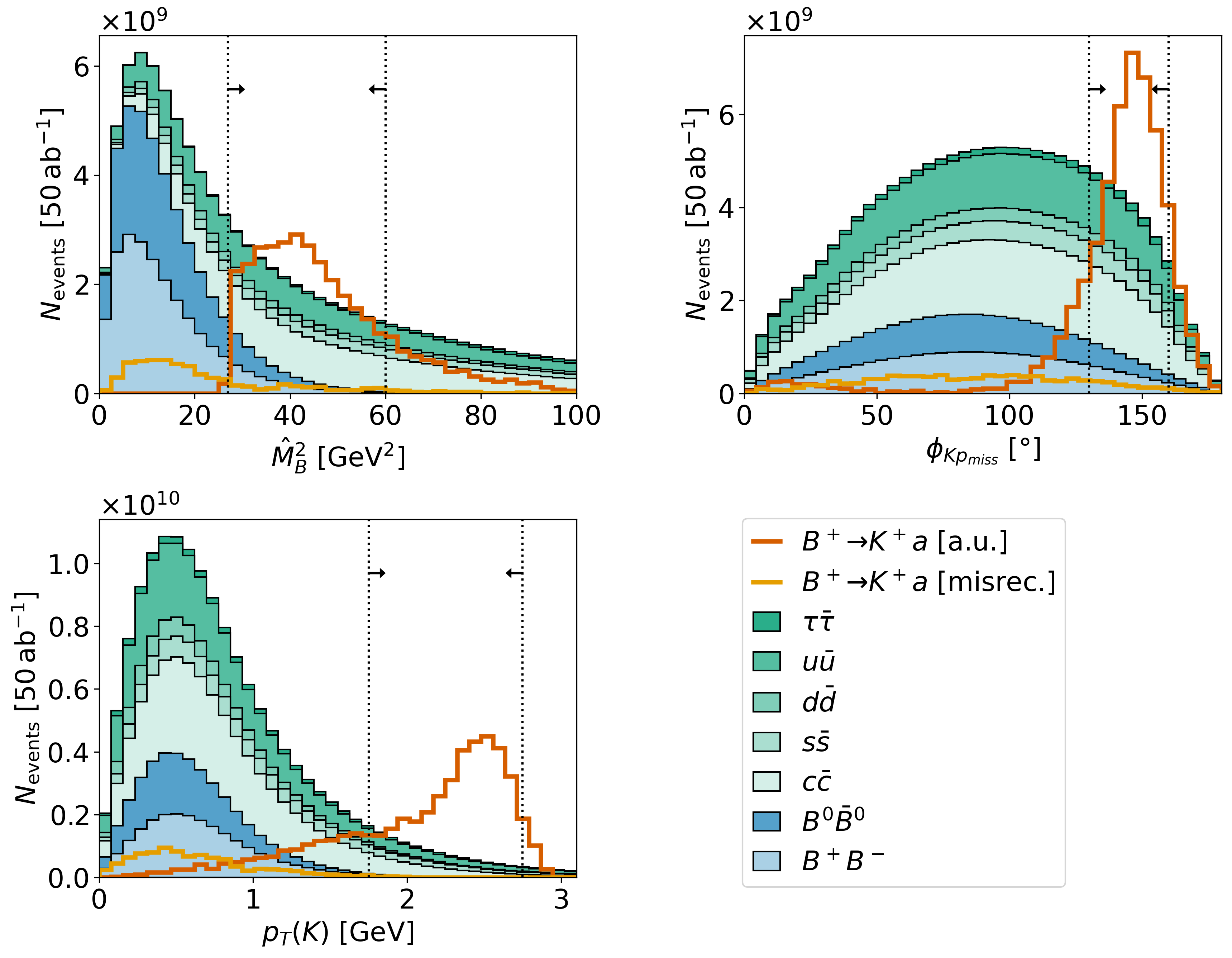}
  \caption{Kinematic distributions of ALPs produced from $B^+\to K^+ a$ for a fixed ALP mass $m_a = 300\,$MeV and various backgrounds before selections. Top left: reconstructed $B$ meson mass, \mb; top right: opening angle of the leading kaon against the missing momentum, \phike; bottom left: transverse momentum of the leading kaon, \ptk. Signal and misreconstructed signal events are normalised arbitrarily to one tenth of the number of background events. Background events are normalised to the total production rates with 50\,ab$^{-1}$ of data luminosity at \belletwo. The dotted vertical lines indicate selection cuts; the arrows point toward the signal region. Figure~\ref{fig:stacked_post} shows the same distributions after selections. 
    \label{fig:stacked_pre}}
\end{figure}

\begin{figure}[t!]
  \centering
  \includegraphics[width=\textwidth]{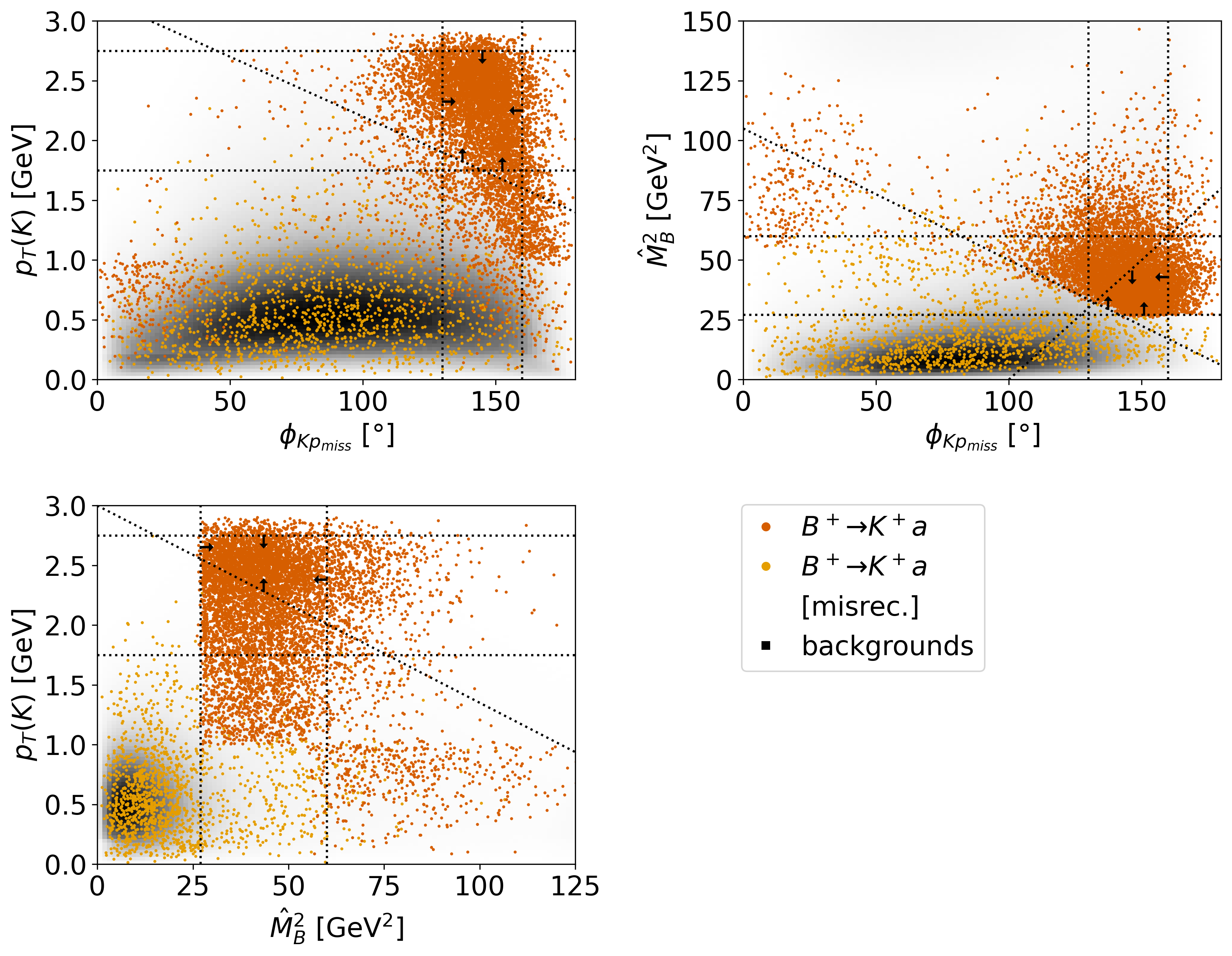}
  \caption{Two-dimensional distributions of the signal (orange points), misreconstructed signal (yellow points), and background (black histograms). Shown are projections of the three-dimensional space of the kinematic variables \ptk, \phike, and \mb. The number of background events corresponds to the full simulated sample of 10\,M events per channel. While the shown background distribution looks well-localised, the tails of the distribution extend into most of the shown parameter space, but at much smaller rates. Signal events correspond to a sample of 10\,k, which we use here for illustration. The dotted lines indicate selections; the arrows point toward the signal region. Figure~\ref{fig:2d_post} shows the same distributions after selections.
    \label{fig:2d_pre}}
\end{figure}

In Fig.\,\ref{fig:signal-kinematics} we show the kinematic distributions of the signal events before further kinematic selections for four different ALP mass benchmarks. Kinematic distributions for the various backgrounds and a signal with a fixed mass \hbox{$m_a=300\,$MeV} are shown for comparison in Figs.\,\ref{fig:stacked_pre} and \ref{fig:2d_pre}. 
%

We point out several relevant features: The two-body kinematics of the signal $B$ decay results in a peaking \ptk distribution of the leading kaon in the lab frame.
The spectrum for the signal kaon is generally harder than for the background, resulting in a high selection purity.The angular distribution \phike peaks at large angles, as expected from a two-body decay of a slow $B$ meson. The distribution is broadened for heavier ALPs. The reconstruction of \mb is dominated by the tag-side $B$ meson and hence is largely independent of the ALP mass. The tail of the signal \mb distribution comes from neutrinos and other missed particles produced from tag-side $B$ decays. For ALP masses below a few hundred MeV all kinematic distributions look identical to the $m_a = 5\,$MeV benchmark.

For signal events, the variables \ptk and \phike exhibit a rather strong correlation. We also observe a strong, but different correlation of \ptk and \phike for misreconstructed signal events, as well as a strong correlation of \phike and \mb. Background events show strong correlations in all combinations of the three variable. We exploit the different correlations for signal and background events and the characteristic shapes of the distributions to optimize our event selection.

\begin{figure}[t]
  \centering
  \includegraphics[width=\textwidth]{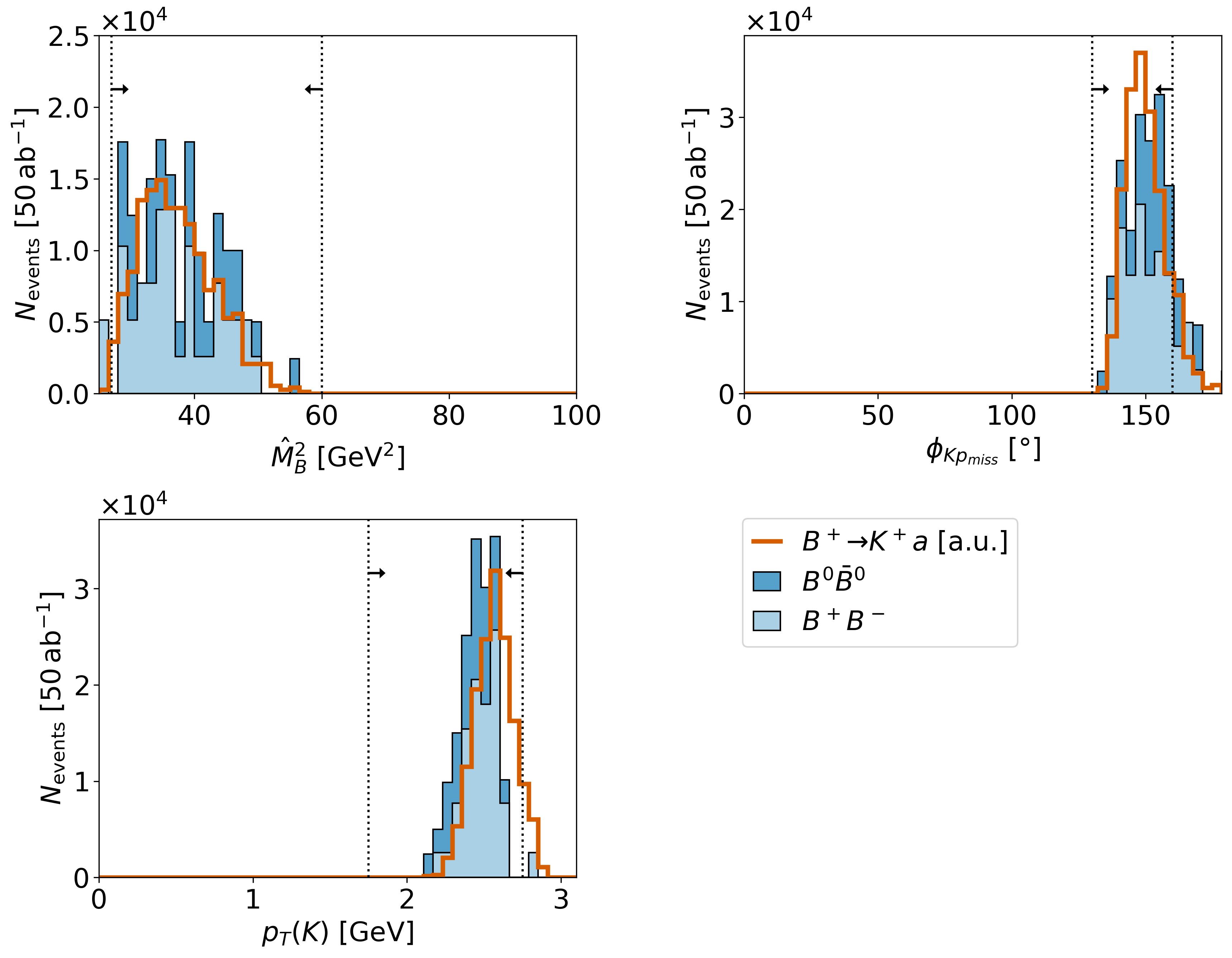}
  \caption{Kinematic distributions of ALPs produced from \bka for a fixed ALP mass $m_a = 300\,$MeV (orange) and various backgrounds (blue) after selections. Selection cuts have been applied to the respective two variables not shown in the distribution. The dotted lines and arrows denote these selections applied in the direction of the errors. No misreconstructed signal and continuum background events in our respective samples are left after cuts. Signal events are normalised arbitrarily to five times the number of background events. Background events are normalised to the total production rates with 50\,ab$^{-1}$ of data luminosity at \belletwo. Figure~\ref{fig:stacked_pre} shows the same distributions after selections. \label{fig:stacked_post}}
\end{figure}

\begin{figure}[t!]
  \centering
  \includegraphics[width=\textwidth]{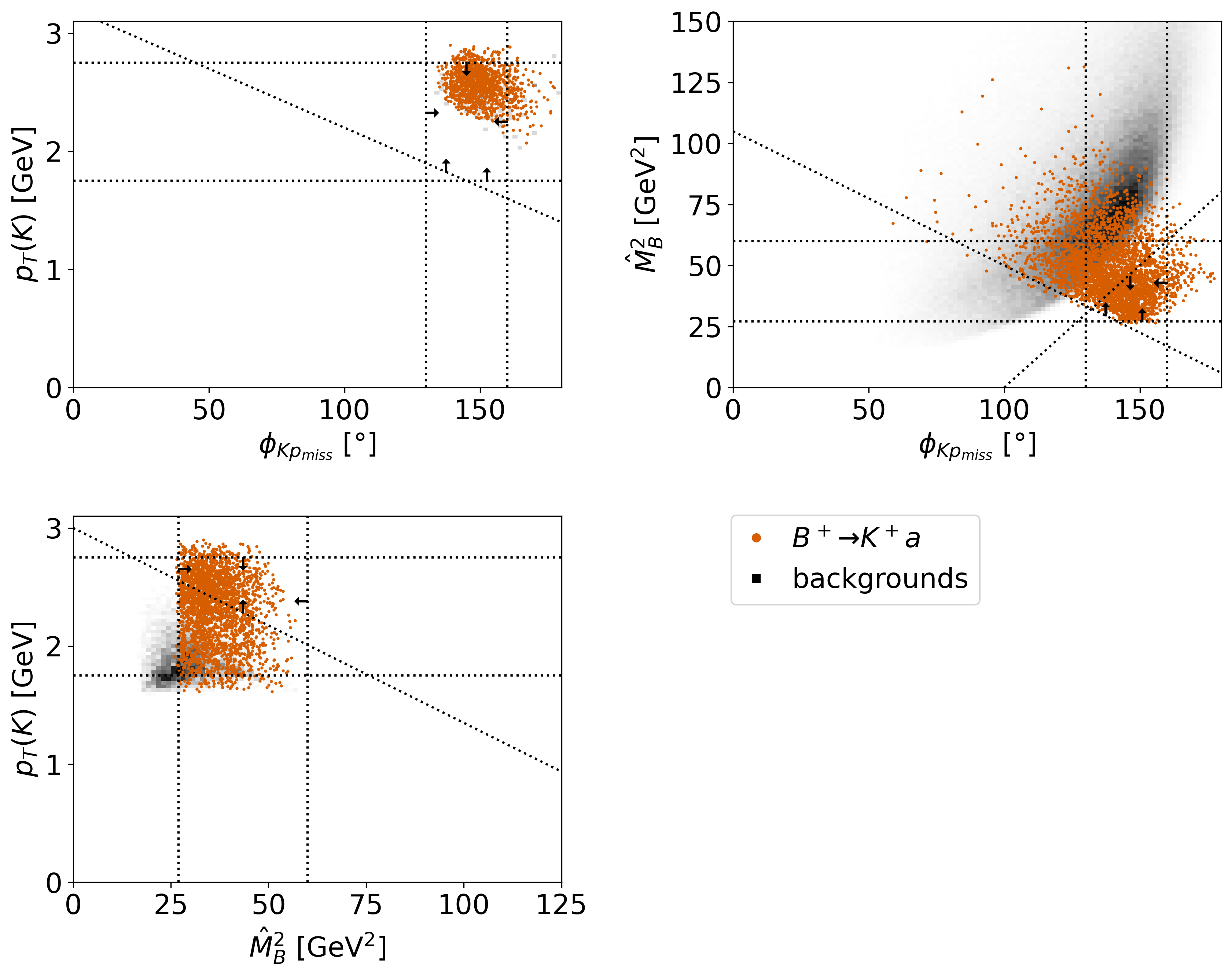}
  \caption{Two-dimensional phase-space distributions of the signal (orange points) and background (black histograms) after selections. Shown are projections of the three-dimensional space of the kinematic variables \ptk, \phike, and \mb. Selection cuts have been applied to the respective third variable not shown in the panel. The dotted lines indicate the selection cuts; the arrows point toward the signal region. No misreconstructed signal events are left after selections. In each plot, the background
  is normalised to the number of events left over after cuts, so that the gray levels correspond to different rates in each plot in this figure and in
  Fig.~\ref{fig:2d_pre}. Signal events correspond to a sample of 10\,k for illustration. Figure~\ref{fig:2d_pre} shows the corresponding distributions before selections. \label{fig:2d_post}}
\end{figure}

\begin{figure}[t!]
    \centering
    \includegraphics[scale=0.5]{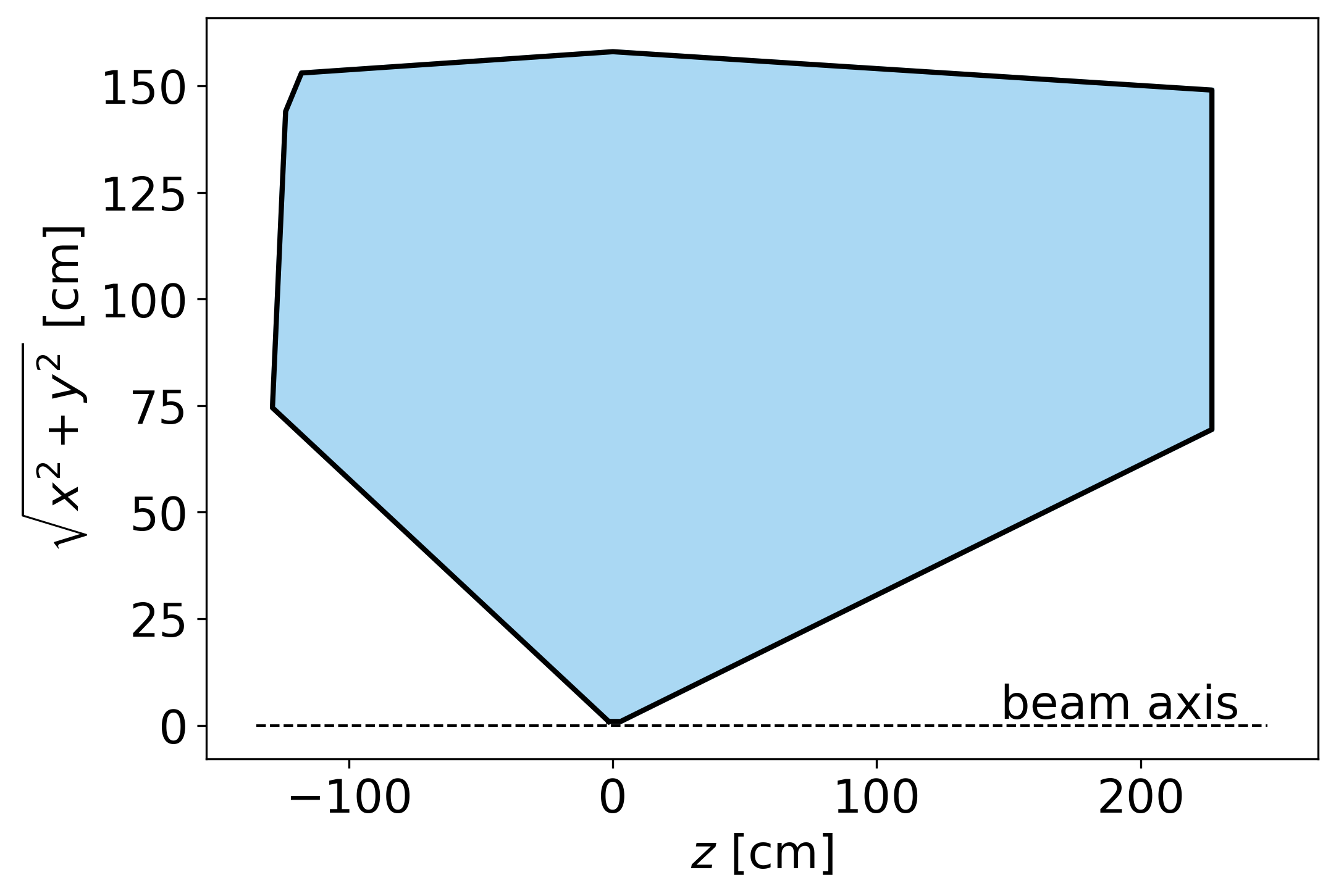}
    \caption{Schematic drawing of the \belletwo detector used to select invisible ALP decays. The displayed geometry includes the central drift chamber and the calorimeter, as specified in the text. The detector is assumed to be symmetric around the beam axis $z$.
    \label{fig:detector}}
\end{figure}

To enhance the signal-to-background ratio, we study ALP mass dependent selection criteria using the aforementioned three variables. All selection criteria are chosen to maximize the Punzi figure of merit for a $5\,\sigma$ discovery~\cite{Punzi:2003bu}. For ALP masses $m_a \lesssim 1\,$GeV, we find that the optimal selections vary little with the ALP mass. We therefore apply one single set of selections to all considered mass benchmarks: 
\begin{align}\label{eq:selection}
    1.75 \leq \ptk &\leq 2.75\,\mathrm{GeV},\\\nonumber
    \ptk &\geq 3.2\,\mathrm{GeV} - 0.01\,\mathrm{GeV}/^\circ \cdot \phike,\\\nonumber
    \ptk &\geq 3.0\,\mathrm{GeV} - 0.0165\,\frac{\mathrm{1}}{\mathrm{GeV}} \cdot \mb,\\\nonumber
    130^\circ \leq \phike &\leq 160^\circ,\\\nonumber
    27 \leq \mb &\leq 60\,\mathrm{GeV^2},\\\nonumber
    \mb  &\geq 105\,\mathrm{GeV^2} - 0.55\,\mathrm{GeV^2}/^\circ \cdot \phike,\\\nonumber
    \mb  &\leq -100\,\mathrm{GeV^2} + 0.992\,\mathrm{GeV^2}/^\circ \cdot \phike.
\end{align}
In Figs.\,\ref{fig:stacked_post} and~\ref{fig:2d_post} we show the kinematic distributions of the signal for $m_a = 300\,$MeV and the remaining backgrounds after applying these selection cuts.
%
 We veto events if the ALP decay vertex is within the geometric acceptance of the tracking detectors or the electromagnetic calorimeter, as illustrated in Fig.\,\ref{fig:detector}.

After applying the selection cuts from~\eqref{eq:selection} and including the detector acceptance, the signal efficiency $\varepsilon = N_{\textrm{sel}}/N_{\textrm{gen}}$ ranges around 10\,\%. Here $N_{\textrm{gen}}$ is the number of generated signal events and $N_{\textrm{sel}}$ is the number of events after selection. In Table\,\ref{tbl:signal}, we show the number of signal and misreconstructed signal events before and after the selection.
\begin{table}[t!]
    \centering
    \begin{tabular}{c|cc|cc}
           & \multicolumn{2}{c|}{before selection} & \multicolumn{2}{c}{after selection}\\
        $m_a$ [GeV] & $N_{\mathrm{signal}}$ & $N_{\mathrm{misrec.}}$  & $N_{\mathrm{signal}}$  & $N_{\mathrm{misrec.}}$  \\\hline
        0.005 & 7802 & 1442 & 1091 & 0\\
        0.3 & 7823 & 1453 & 1022 & 0\\
        1 & 7737 & 1460 & 770 & 0\\
        3 & 7568 & 1649 & 0 & 0
    \end{tabular}
    \caption{Number of signal and misreconstructed signal events before and after applying the selection cuts from \eqref{eq:selection} for various ALP mass benchmarks. The numbers are based on 10k generated signal events. The difference between the number of generated events, $N_{\rm gen}$, and the sum of signal and misreconstructed events, $N_{\rm signal} + N_{\rm misrec.},$ is due to kaons that are out of acceptance or not identified as kaons.
    \label{tbl:signal}}
\end{table}

 After selection, we have 39 \charged and 28 \mixed background events left, while no background from $\tau^+\tau^-$ and continuum events remains in generated samples of 10M events per background source. We finally scale these numbers to the event rates corresponding to the respective integrated luminosity. This procedure potentially underestimates background from $\tau^+\tau^-$ and continuum events due to our limited background statistics. Based on our studies we assume that the dominant background comes from \charged and \mixed events. 

\subsection{Projected sensitivity}
\begin{figure}[t!]
    \centering
    \includegraphics[scale=0.5]{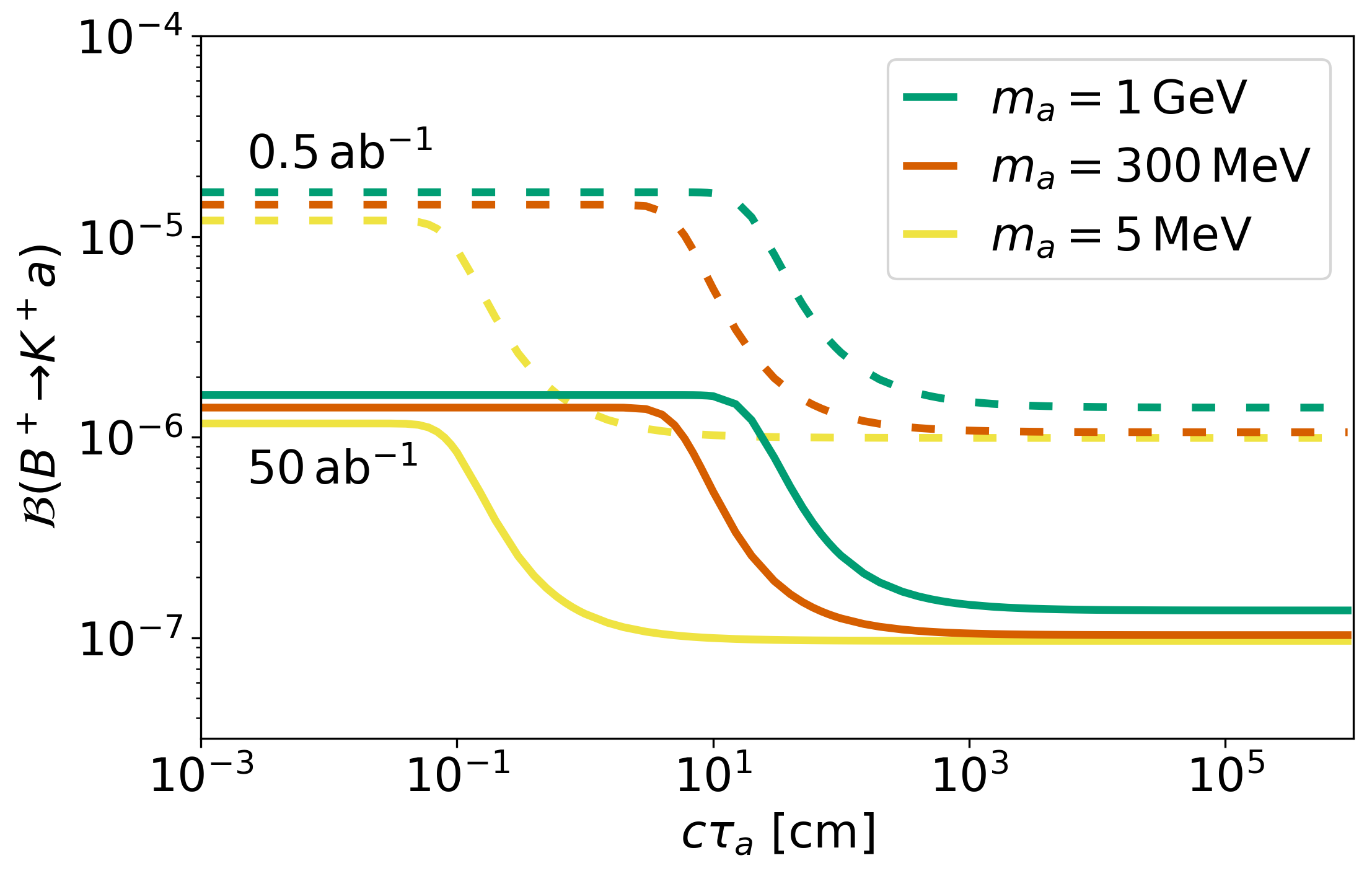}
    \caption{Projected $90\,\%\,$CL upper limits  on the branching ratio $\Br(B^+\to K^+ a)$ of an invisibly decaying ALP or similar (pseudo-)scalar resonance $a$ from $B^+\to K^+a,\,a\to \me$ at \belletwo. The projected limits are shown for fixed ALP masses as a function of the ALP decay length $c\tau_a$, for 0.5\,\invab (dashed) and 50\,\invab (solid) of data.}
    \label{fig:results_Br}
\end{figure}

\begin{figure}[t!]
  \centering
  \includegraphics[width=\textwidth]{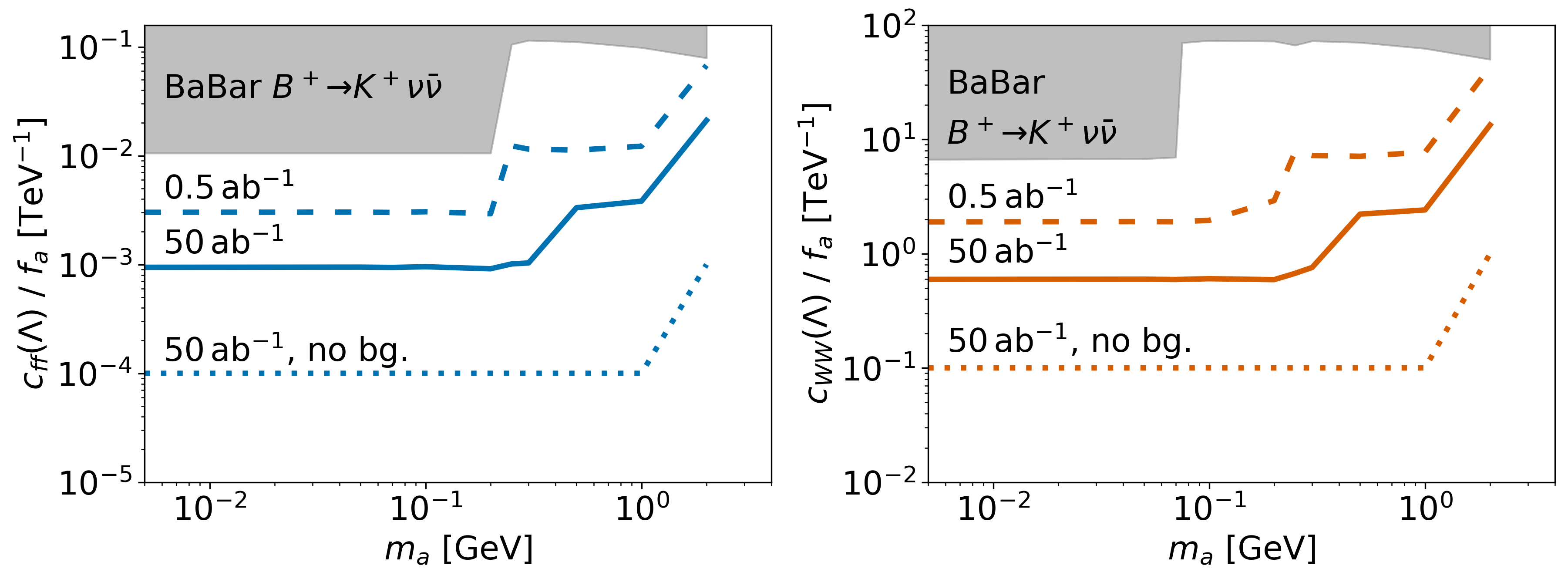}
  \caption{Projected 90\,\%-CL upper limits on the couplings $c_{ff}$ (left) and $c_{WW}$ (right) in our two scenarios. The solid lines denote the full \belletwo luminosity of 50\,\invab and dashed lines denote the current integrated luminosity 0.5\,\invab, both with the full background considered. The dotted line shows the limit with full 50\,\invab luminosity for the case that the background is reduced to zero. The shaded regions show the \babar bound from $B^+\to K^+\nu\bar{\nu}$ for comparison. \label{fig:results}}
\end{figure}

For each ALP mass scenario, we derive an expected 90\% CL upper limit on the observed number of signal events, $N_S$, based on the expected number of background events, $N_B$. Here $N_B$ is the number of simulated background events after our selection procedure, scaled to the respective integrated luminosity. In each ALP mass scenario, the  90\,\% CL upper limit on the signal rate, $N_S$, is determined iteratively, such that the Poisson probability of observing $N$ events when predicting $N_S + N_B$ events is 0.1. Here $N$ is the integer closest to the number of expected background events $N_B$. In this way, we derive upper bounds on the branching ratio $\mathcal{B}(\bka)$ by requiring
 \begin{align}
    N_S \ge \nbb\cdot \mathcal{B}(\bka)\cdot \langle\mathds{P}_a\rangle.
\end{align}
The projected upper limits on $\Br(B^+\to K^+ a)$ are shown as a function of the ALP decay length $c\tau_a$ in Fig.\,\ref{fig:results_Br}. We interpret these bounds in terms of the couplings $c_{ff}$ and $c_{WW}$; the results are shown in Fig.\,\ref{fig:results}.
 To obtain the bound as a function of the ALP mass and couplings, we use the root-finding algorithm \texttt{scipy.optimize.fsolve}~\cite{2020SciPy-NMeth} and show an additional, optimistic, projected limit for the case that the background is reduced to zero. All curves feature a mass-dependent kink. For ALP masses below this kink, the ALP can be considered stable at the scale of the \belletwo vertex detector. For ALP masses above the kink, the decay length of the ALP is shorter than the detector radius, but the ALP can still appear invisible due to the limited acceptance and reconstruction efficiency. In this way, searches for $B^+\to K^+ \me$ are sensitive to even short-lived light resonances, which have a statistical probability to decay close to the production point, but outside the angular acceptance of the detector.
 
 The sensitivity to heavy, short-lived ALPs in Fig.\,\ref{fig:results_Br} is thus due to the detector geometry and applies for all ALPs that do not decay to neutrinos and/or invisible new particles.
 
\begin{figure}[th!]
  \centering
  \includegraphics[scale=0.47]{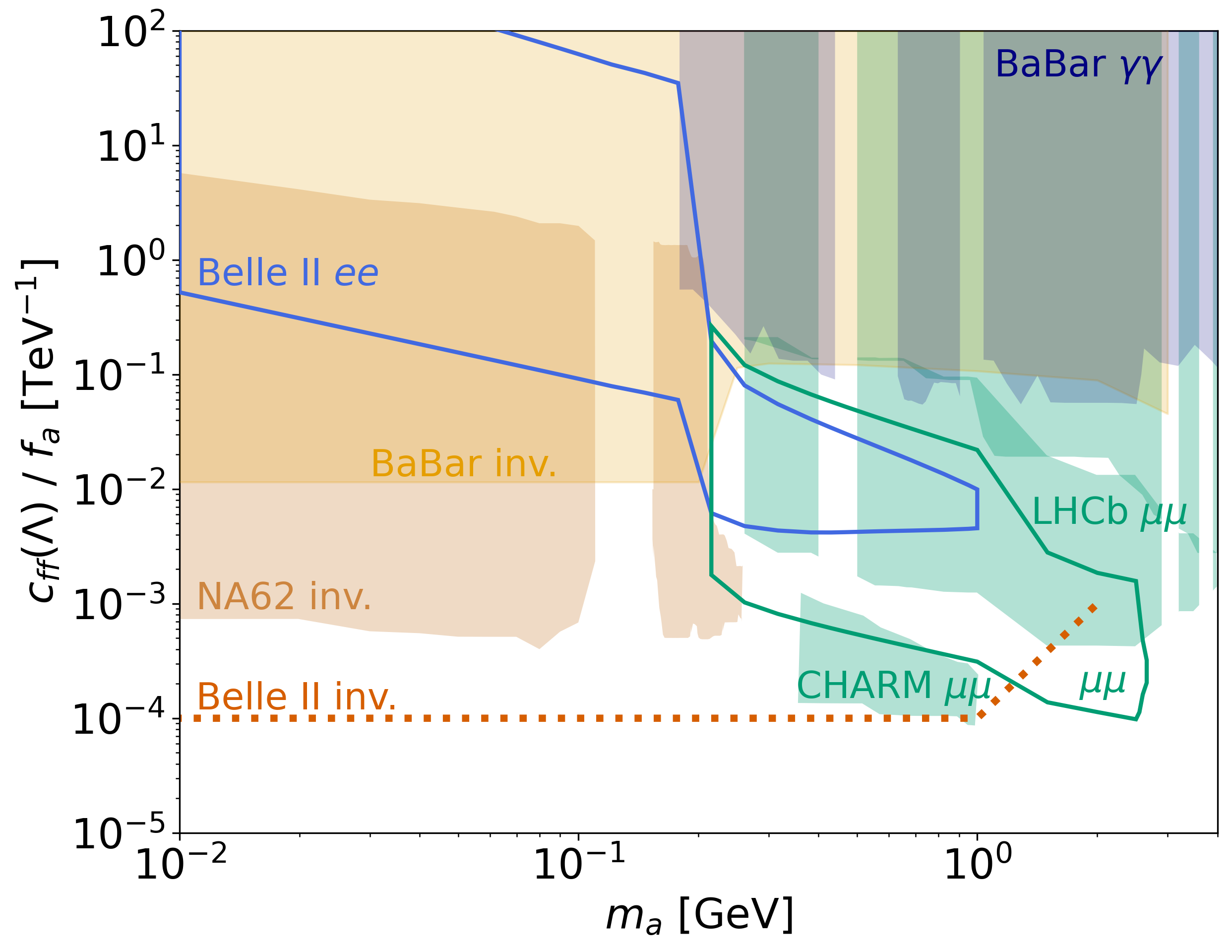}\\\vspace*{0.7cm} 
  \includegraphics[scale=0.47]{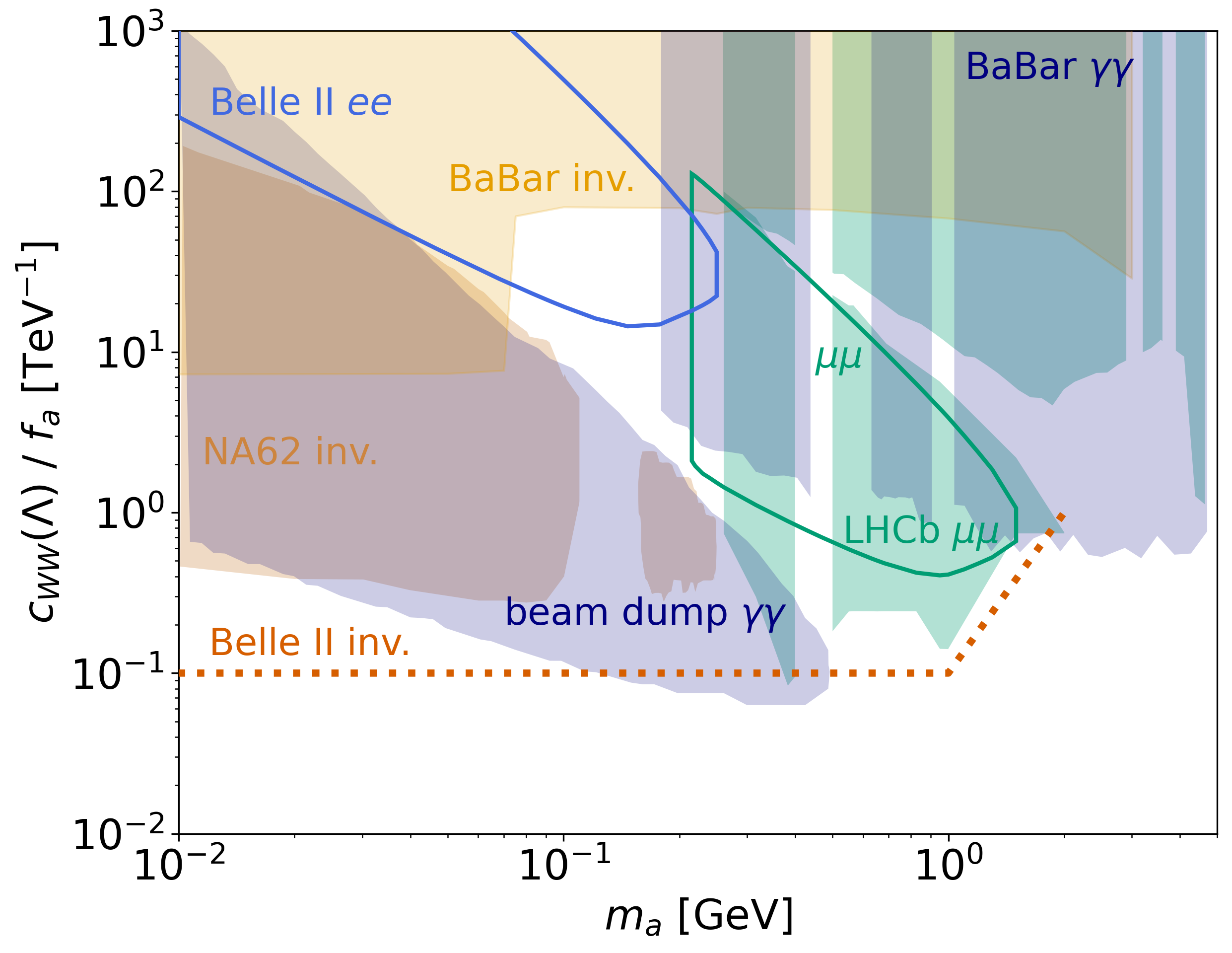}
  \caption{Projections for \belletwo's sensitivity to ALPs with effective couplings to fermions, $c_{ff}(\Lambda)/f_a$ (top), and weak gauge bosons, $c_{WW}(\Lambda)/f_a$ (bottom), as a function of the ALP mass $m_a$. Shown are contours of $N=2.3$ events expected in $50\,$ab$^{-1}$ of data from $B^+ \rightarrow K^+ a,\ a \rightarrow X$ decays with $X= e^+ e^-$ (blue) and $\mu^+ \mu^-$ (green). The area enclosed by the contours could be excluded at the $90\%$ CL, assuming zero background. The region above the orange dashed curve (cf. Fig.~\ref{fig:results}) could be excluded at $90\%$ CL with the search for $B^+\to K^+ a,\ a\to$ inv. proposed in Sec.~\ref{SEC:invisible}, assuming zero background. For comparison, we also show the existing bounds on ALPs from Fig.~\ref{fig:bounds}.\label{fig:projections_combined}}
\end{figure}

 We finally show our projected sensitivity for $B^+\to K^+\me$ together with existing bounds from other searches in Fig.\,\ref{fig:projections_combined}.
 
\section{Displaced ALP decays at Belle II}
\label{SEC:displaced}
As discussed in Sec.~\ref{SEC:bounds}, current bounds show a complementarity of invisible and displaced searches when scanning the ALP parameter space, see~Fig.~\ref{fig:bounds}. Indeed, for a fixed coupling, light ALPs are produced with a large boost and tend to decay outside of a detector, leaving signatures with missing momentum. Heavier ALPs, on the other hand, are more likely to decay within the detector volume and can be detected through displaced decay products.

To complement our analysis of invisible ALPs at \belletwo from Sec.~\ref{SEC:invisible}, we assess the reach of displaced signatures at \belletwo. We build our analysis on a recent study of dark scalars $S$ in $B^+ \rightarrow K^+ S,\ S \rightarrow X$~\cite{Filimonova:2019tuy} for leptonic final states $X=\{ e^+ e^-,\, \mu^+ \mu^-\}$. Since both the dark scalar and the ALP are bosons with spin zero, their kinematic distributions in the relevant decay chains are the same. This allows us to rerun the analysis for ALPs, using the production rate, lifetime and decay branching ratios from Sec.~\ref{SEC:alps}. Following the procedure of Sec.~\RomanNumeralCaps{4} in Ref.~\cite{Filimonova:2019tuy}, we calculate the number of displaced lepton pairs from ALP decays within the \belletwo tracking detector for the two ALP scenarios.
 
In Fig.~\ref{fig:projections_combined}, we show contours of 2.3 expected signal events from $B^+ \rightarrow K^+ a,\ a \rightarrow \{e^+e^-,\,\mu^+\mu^-\}$ decays at \belletwo in 50\,ab$^{-1}$ of data. The regions inside the contours would be excluded at the 90\% CL by a non-observation of ALP decays, assuming zero background. For a given ALP mass, the sensitivity to larger couplings is limited by the minimum radial displacement required for the ALP decay products to be $d>0.9\,$cm.  The sensitivity to small couplings is limited by the ALP production rate. In the $c_{ff}$ scenario, the sudden changes in sensitivity at $m_a \approx 0.2\,$GeV and $1\,$GeV are due to changes in the ALP branching ratios, see~Fig.~\ref{fig:branching ratios}. The sensitivity cutoff at $m_a\sim1$~GeV for electrons and $m_a\sim2.5$~GeV for muons are also largely due to drops in the ALP branching ratios. In the $c_{ff}$ scenario the sensitivity to displaced leptons is expected to dominate due to the large branching ratio into leptons; for the $c_{WW}$ scenario the sensitivity to photons is more promising. Searches for $B^+\to K^+a, a\to \gamma\gamma$ at \belletwo can be interesting probes of long-lived ALPs with dominant electroweak couplings. A recent such search by BaBar~\cite{BaBar:2021ich}, shown in blue in Fig.~\ref{fig:projections_combined}, suggests that \belletwo has good potential to explore photon final states. For a realistic prediction of the reach, however, the reconstruction efficiency for photon pairs and the background in displaced di-photon searches have to be assessed.

To compare the reach of displaced and invisible signatures at \belletwo, we also show the projections for $B^+\to K^+ a,\ a\to\,$inv. from Fig.~\ref{fig:results} for zero background (orange dotted curves). For a fixed coupling, displaced searches are most sensitive at high ALP masses, while invisible searches appear to be much more sensitive to lighter ALPs with $m_a<1\,$GeV. For a more accurate comparison of the two searches, a detailed study of backgrounds for the displaced ALP signatures would be required.
 
In combination, displaced and invisible signatures at \belletwo can set mass-dependent upper bounds on the ALP couplings
\begin{align}
\frac{c_{ff}(\Lambda)}{f_a} \lesssim \frac{10^{-4}}{\text{TeV}},\qquad \frac{c_{WW}(\Lambda)}{f_a} \lesssim \frac{10^{-1}}{\text{TeV}}\,,
\end{align}
improving the current reach by up to two orders of magnitude, see \eqref{eq:current-bounds}, and covering unexplored parameter space at larger couplings. Due to the high projected sensitivity to $B^+\to K^+ a,\ a\to$ inv. even at moderate ALP lifetimes, signatures with missing energy play an important role in probing such feeble interactions. In the case of an observed excess in $B\to K \me$ decays, searches for displaced visible ALP decays will provide a valuable independent test of the underlying model.

\section{Conclusions}
\label{SEC:conclusions}
The purpose of this paper was to assess \belletwo's sensitivity to displaced versus invisibly decaying light resonances produced from meson decays. To this end, we have developed a new search strategy for invisibly decaying ALPs in meson decays $B^+\to K^+ a,\, a\to \me$.

The search for $B^+\to K^+ \me$ is optimized for two-body decays and is inclusive in the decay modes of the second $B$ meson from $B\bar{B}$ pairs produced at \belletwo. We have performed a detailed analysis of the signal and background kinematics and identified three variables as very sensitive to $B^+\to K^+ \me$: the reconstructed $B$ meson mass squared, $\mb$; the transverse momentum of the kaon, $\ptk$; and the angle between the kaon and the missing momentum, $\phike$, the latter being reconstructed indirectly from the visible decay products. With suitable selections based on three kinematic variables, we were able to determine a signal region that is largely free from SM background and misreconstructed \bka decays.

For ALPs with decay lengths $c\tau_a \gtrsim 1\,\text{cm}$, we find that \belletwo can probe rates of $\Br(\bka) \gtrsim 10^{-7}$ with $50\,$ab$^{-1}$ of data. Remarkably, the search is also sensitive to short-lived ALPs, which might escape the detector due to the limited angular coverage. For ALPs with decay lengths $c\tau_a \lesssim 1\,\text{cm}$, we expect \belletwo to be sensitive to $\Br(B^+\to K^+ a) \gtrsim 10^{-6}$ with $50\,$ab$^{-1}$. The sensitivity can be further enhanced with a multi-variate analysis. With the $0.5\,$ab$^{-1}$ of data expected in 2022, \belletwo can already probe ALPs with smaller production rates than our reinterpretation of a search for $B^+\to K^+\nu\bar{\nu}$ by BaBar.

To compare our predictions for $B^+\to K^+\me$ with searches for displaced particles at \belletwo, we have estimated the reach of $B^+\to K^+ a,\, a\to X$ with $X=\{e^+e^-,\mu^+\mu^-\}$ for two specific scenarios with ALPs coupling to fermions and weak gauge bosons, respectively. For ALPs with masses below about 1\,GeV, the search for invisible decays $B^+\to K^+\me$ can probe couplings up to several orders of magnitude smaller than searches for displaced decays in $B^+\to K^+ X$. For heavier ALPs, we expect that searches for displaced decays can be more sensitive.

Compared with existing collider searches for displaced and invisible light resonances, \belletwo can significantly improve the sensitivity to small ALP couplings, see Fig.~\ref{fig:projections_combined}. For sub-GeV ALPs, the sensitivity of $B^+\to K^+\me$ even exceeds the reach of long-baseline experiments like NA62 and CHARM and is competitive with high-intensity beam-dump experiments like NuCal and E137.

For ALPs with predominant couplings to photons, we expect a significantly improved sensitivity also from approved experiments like FASER, NA62-dump, and NA64e in the near future~\cite{Agrawal:2021dbo}. An even better sensitivity may be reached by the proposed experiments SHiP, FASER2, or ultimately at a Gamma Factory in the far future. For ALPs with predominant couplings to fermions, the proposed beam-dump experiment SHADOWS\,\cite{Baldini:2021hfw} might reach the best sensitivity, while ultimate sensitivity can be reached by the proposed experiments KLEVER (low mass) and MATHUSLA (high mass)~\cite{Agrawal:2021dbo}.

Our predictions for $B^+\to K^+\me$ and $B^+\to K^+ X$ at \belletwo apply in a very similar way to other light pseudo-scalar or scalar resonances. \belletwo has just started to exploit its discovery potential for light new particles with signatures of displaced vertices or missing energy. We hope that this analysis will be used as a guideline for future searches.

\acknowledgments
We thank Sebastian Bruggisser and Lara Grabitz for providing us with their code for the RG evolution of the ALP couplings and Christopher Smith and Andrea Thamm for discussions. We also thank Tobioka Kohsaku, Sophie Renner, Robert Ziegler and Jure Zupan for discussions and for spotting a mistake in our interpretation of the NA62 results. The research of AF is supported by the NWO Vidi grant ``Self-interacting asymmetric dark matter". RS acknowledges support of the \emph{Deutsche Forschungsgemeinschaft} (DFG) through the research training group \emph{Particle Physics Beyond the Standard Model} (GRK 1940). The research of SW is supported by the DFG under grant no. 396021762–TRR 257. TF is supported by the Helmholtz (HGF) Young Investigators Group grant no.\ VH-NG-1303.

\clearpage

\appendix
\section{ALP decay widths}
\label{app:decay-widths}
For convenience, we summarize the partial decay widths of the ALP in terms of its mass $m_a$ and couplings $c_{ii}$. All couplings are defined at the ALP mass scale, $c_{ii} = c_{ii}(m_a)$.
 In large parts, the formulas below correspond with the results of Ref.~\cite{Bauer:2020jbp}.
 
 The ALP decay widths into fermions is given by
\begin{align}
    \Gamma_{a\to f\bar{f}} &= 2\pi m_a N_c^f \frac{\left|c_{ff}(m_a)\right|^2 m_f^2}{\Lambda^2} \left(1-\frac{4m_f^2}{m_a^2}\right)^{\frac{1}{2}} + \mathcal{O}\left(\frac{\alpha^2}{(4\pi)^2}\, c_{WW}\right), \label{eq:atofermions}\quad f = \{q,\ell\}
\end{align}
where $N_c^\ell=1$, $N_c^q=3$.
The decay width into hadrons reads
\begin{align}
    \Gamma_{a\to\text{had}} &= \frac{2\alpha_s^2m_a^3}{\pi}\frac{\left|C_{GG}^{\text{eff}}(m_a)\right|^2}{\Lambda^2}\left(1+\left(\frac{97}{4}-\frac{7n_q}{6}\right)\frac{\alpha_s}{\pi}\right) + \sum_{q}\Gamma_{a\to q\bar{q}}\,, \label{eq:atohadrons}
 \end{align}
 where $n_q=3$ is the number of light quarks $q=\{u,d,s\}$
 and the effective gluon coupling is given by
 \begin{align}
     C_{GG}^{\text{eff}}(m_a) &= c_{GG}(m_a) + \sum_{q'} \frac{c_{q'q'}(m_a)}{2} B_1\!\left(\frac{4m_{q'}^2}{m_a^2}\right) \label{eq:cGGeff}
 \end{align}
 where $q'$ are all quarks with $m_{q'} \lesssim m_a$.
We use $\Gamma_{a\to \text{had}}$ for ALP masses above 1\,GeV. For $m_a < 1\,$GeV, we approximate the  hadronic decay width of the ALP by the decay width into three pions,
 \begin{align}
    \Gamma_{a\to\pi^0\pi^i\pi^j} &= \frac{m_am_\pi^4}{384\pi f_\pi^2\Lambda^2} \left(c_{uu}(m_a)-c_{dd}(m_a)+2c_{GG}(m_a)\frac{m_d-m_u}{m_d+m_u}\right)^2 g_{ij}\left(\frac{m_\pi^2}{m_a^2}\right), \label{eq:atopions}
\end{align}
 with
 \begin{align}
    g_{00}(r) &= \frac{2}{(1-r)^2} \int_{4r}^{(1-\sqrt{r})^2} \!\!\!\!\!\!\!\!\!\!\!\!\!\!\!\!\!\mathrm{d}z \:\:\:\:\: \sqrt{1-4\frac{r}{z}}\sqrt{\lambda(1,z,r)}\label{eq:g00pions}\\\nonumber
    g_{+-}(r) &= \frac{12}{(1-r)^2} \int_{4r}^{(1-\sqrt{r})^2} \!\!\!\!\!\!\!\!\!\!\!\!\!\!\!\!\!\mathrm{d}z \:\:\:\:\: \sqrt{1-4\frac{r}{z}}\sqrt{\lambda(1,z,r)}\: (z-r)^2
 \end{align}
 and $\lambda(a,b,c)=a^2+b^2+c^2 -2ab -2 bc -2ca$.
 
 The decay width of the ALP into photons is given by
\begin{align}
    \Gamma_{a\to\gamma\gamma} &= \frac{\alpha^2m_a^3}{4\pi\Lambda^2} \left|C_{\gamma\gamma}^{\text{eff}}(m_a)\right|^2, \label{eq:atophotons}
\end{align}
with
\begin{align}
    C_{\gamma\gamma}^{\text{eff}}(m_a) &= \left\{\begin{array}{ll}c_{\gamma\gamma}(m_a) + \sum\limits_{f\in\{\ell,Q\}} N_c^fQ_f^2\,c_{ff}(m_a)B_1\!\left(\frac{4m_f^2}{m_a^2}\right) & m_a>1\,\text{GeV}\\
    \begin{array}{l}c_{\gamma\gamma}(m_a) + \sum\limits_{f\in\{\ell,Q\}} N_c^fQ_f^2\,c_{ff}(m_a)B_1\!\left(\frac{4m_f^2}{m_a^2}\right)\\ \phantom{=}- \frac{m_a^2}{m_\pi^2-m_a^2} \frac{c_{uu}(m_a)-c_{dd}(m_a)}{2} - \left(\frac{5}{3}+\frac{m_\pi^2}{m_\pi^2-m_a^2}\frac{m_d-m_u}{m_d+m_u}\right)c_{GG}(m_a)\end{array} & m_a<1\,\text{GeV},\end{array}\right. \label{eq:cgamgameff}
\end{align}
where $\ell\in\{e,\mu,\tau\}$, $Q\in\{c,b,t\}$ and
\begin{align}
    B_1(\tau) &= 1-\tau f^2(\tau) \label{eq:B1}\\
    f(\tau) &= \left\{\begin{array}{ll}\arcsin\frac{1}{\sqrt{\tau}} & \tau\geq1\\\nonumber
    \frac{\pi}{2}+\frac{i}{2}\ln\frac{1+\sqrt{1-\tau}}{1-\sqrt{1-\tau}} & \tau<1\end{array}\right.\label{eq:f}\,.
\end{align}

\bibliographystyle{JHEP_improved}
\bibliography{alps_belle2}

\end{document}